\documentclass[11pt]{article}
\pdfoutput=1
\newcommand{\Comment}[1]{{}}
\usepackage{amsfonts,amsthm,amsmath,amssymb,slashed}
\usepackage[textwidth = 430 pt, textheight = 630 pt]{geometry}
\usepackage{color}
\usepackage{booktabs}
\usepackage{subcaption}
\usepackage[export]{adjustbox}

\Comment{\usepackage{color}
\definecolor{MyDarkBlue}{rgb}{0.15,0.15,0.45}
\usepackage[linktocpage=true]{hyperref}
\hypersetup{
colorlinks=true,
citecolor=MyDarkBlue,
linkcolor=MyDarkBlue,
urlcolor=MyDarkBlue,
pdfauthor={Luis Alejo and Horatiu Nastase},
pdftitle={Particle-vortex duality and theta terms in AdS/CMT applications},
pdfsubject={hep-th}
}
\usepackage[utf8]{inputenc}
\usepackage[numbers,sort&compress]{natbib}
\usepackage{hypernat}}
\usepackage{graphicx}
\usepackage{cite}
\usepackage[colorlinks=true]{hyperref}
\hypersetup{
  allcolors = blue,
}

\newcommand\ignore[1]{}
\def\one{{\,\hbox{1\kern-.8mm l}}}

\def\a{\alpha}\def\b{\beta}

\def\d{\partial}

\newcommand{\Cset}{{\,\,{{{^{_{\pmb{\mid}}}}\kern-.45em{\mathrm C}}}}}

\newcommand{\be}{\begin{equation}}
\newcommand{\bea}{\begin{eqnarray}}

\newcommand{\ee}{\end{equation}}
\newcommand{\eea}{\end{eqnarray}}

\begin{document}

\renewcommand{\thefootnote}{\fnsymbol{footnote}}

\makeatletter
\@addtoreset{equation}{section}
\makeatother
\renewcommand{\theequation}{\thesection.\arabic{equation}}

\rightline{}
\rightline{}




\begin{center}
{\LARGE \bf{\sc Particle-vortex duality and theta terms in AdS/CMT applications}}
\end{center}
 \vspace{1truecm}
\thispagestyle{empty} \centerline{
{\large \bf {\sc Luis Alejo${}^{a}$}}\footnote{E-mail address: \Comment{\href{mailto:luis.alejo@unesp.br}}{\tt luis.alejo@unesp.br}}
{\bf{\sc and}}
{\large \bf {\sc Horatiu Nastase${}^{a}$}}\footnote{E-mail address: \Comment{\href{mailto:horatiu.nastase@unesp.br}}{\tt horatiu.nastase@unesp.br}}
                                                        }

\vspace{.5cm}


\centerline{{\it ${}^a$Instituto de F\'{i}sica Te\'{o}rica, UNESP-Universidade Estadual Paulista}}
\centerline{{\it R. Dr. Bento T. Ferraz 271, Bl. II, Sao Paulo 01140-070, SP, Brazil}}

\vspace{1truecm}

\thispagestyle{empty}

\centerline{\sc Abstract}

\vspace{.4truecm}

\begin{center}
\begin{minipage}[c]{380pt}
{\noindent In this paper we study particle-vortex duality and the effect of theta terms from the point of view of 
AdS/CMT constructions. We can construct the duality in 2+1 dimensional field theories with or without a Chern-Simons term, and 
derive an effect on conductivities, when the action is viewed as a response action. We can find its effect on 3+1 
dimensional theories, with or without a theta term, 
coupled to gravity in asymptotically AdS space, and derive the resulting effect on conductivities 
defined in the spirit of AdS/CFT. AdS/CFT then relates the 2+1 dimensional and the 3+1 dimensional cases naturally.
Quantum gravity corrections, as well as more general effective actions for the abelian vector, can be treated similarly.
We can use the fluid/gravity correspondence, and the membrane paradigm, to define shear and bulk viscosities 
$\eta$ and $\zeta$ for a gravity plus abelian vector plus scalar system near a black hole, and define the effect of the 
S-duality on it.
}
\end{minipage}
\end{center}

\vspace{.5cm}

\setcounter{page}{0}
\setcounter{tocdepth}{2}

\newpage

\renewcommand{\thefootnote}{\arabic{footnote}}
\setcounter{footnote}{0}

\linespread{1.1}
\parskip 4pt



\section{Introduction}

Particle-vortex duality is a very useful tool in 2+1 dimensional quantum field theories \cite{Zee:2003mt}
(see for instance \cite{Murugan:2016zal,Karch:2016sxi,Seiberg:2016gmd}), though it was still not as used as the 
particle-monopole, or S-duality, in 3+1 dimensional quantum field theories. The duality goes back to work on 
superconductivity \cite{Dasgupta:1981zz,Peskin:1977kp} and anyon superconductivity and the fractional quantum Hall effect \cite{Lee:1989fw} (see also the early works \cite{Marino:1987tk,Marino:1992uu}), and was defined at the level 
of the path integral in \cite{Burgess:2000kj,Murugan:2014sfa} (see also \cite{Ramos:2005yy,Ramos:2007hk} for another 
take on a path integral formulation). 

Within the context of AdS/CFT correspondence \cite{Maldacena:1997re} (see the books 
\cite{Nastase:2015wjb,Ammon:2015wua} for reviews; see also \cite{Nastase:2018cfe}), Witten \cite{Witten:2003ya} 
first described the effect of particle-vortex duality on  current correlators in conformal field theories with abelian symmetries
and speculated on the gravity dual, and afterwards the effect on M theory membranes was sketched in \cite{Herzog:2007ij}, 
but that was before the correct M theory description in terms of the ABJM model \cite{Aharony:2008ug} was developed. 
More precisely, it was shown in \cite{Murugan:2014sfa} in a simple model based on (a reduction of) the ABJM model
that the particle-vortex duality at the level of the path integral in 2+1 dimensional field theory corresponds to 
usual S-duality (or Maxwell duality) in the  3+1 dimensional gravitational bulk dual to it.\footnote{See e.g.,
\cite{LeGuillou:1996dv,LeGuillou:1997zx} for an early example of 2+1 dimensional duality in the path integral.}

We are however interested in understanding better the effects of this particle-vortex duality on the gravity duals relevant 
for condensed matter, i.e., AdS/CMT (see the book \cite{Nastase:2017cxp} for a review), and in particular to transport 
coefficients like conductivity and shear viscosity, evaluated either in the field theory, or in the gravitational bulk. This 
will be the subject of this paper.

The AdS/CFT calculations using the membrane paradigm will be based on the original work of Iqbal and Liu \cite{Iqbal:2008}
(see also \cite{Lopez-Arcos:2013uga}), as well as Kovtun, Son and Starinets \cite{Kovtun:2003wp}
We we also treat quantum corrections to the gravitational action and the Maxwell and scalar fields in it, following the 
formalism of \cite{Brigante:2007nu,Myers:2010pk}. Within this context, the formalism for computing conductivities 
is based on \cite{Banks:2015wha,Donos:2017mhp}. We will also consider the membrane paradigm for the fluid/gravity 
correspondence defined in \cite{Fischler:2015kro} (an early formulation is \cite{Parikh:1997ma}).

The paper is organized as follows. In section 2 we describe particle-vortex duality in the presence of $\Theta$ terms in 
2+1 dimensional field theories, and their effect on conductivities. In section 3 we consider an AdS/CMT set-up for the 
same, with a gravitational theory, with an asymptotically AdS black hole solution with a horizon, in 3+1 dimensions, and 
consider the effect of S-duality on Maxwell fields, and horizon conductivities in a membrane paradigm calculation. 
In section 4 we show that the result is consistent with AdS/CFT, by doing an expansion of the duality relations away 
from either the boundary, or the horizon of the black hole. In section 5 we consider the effect of gravitational 
quantum corrections on the S-duality and resulting conductivities. In section 6 we consider the fluid/gravity correspondence in 
a membrane paradigm calculation, and show that the shear and bulk viscosities are unaffected by matter, thus 
are unaffected by S-duality. In section 7 we conclude, and in the Appendix we review the standard particle-vortex 
duality in 2+1 dimensions.

\section{Particle-vortex duality  in general 2+1 dimensional field theories}

In a very simplified way, particle-vortex duality can be thought of as Poincar\'{e} duality in 2+1 dimensions, exchanging 
a real scalar $\theta$, identified with the angular variable of a (gauged) complex scalar field $\phi$, with a gauge field 
$a_\mu$, by
\be
v(\d_\mu \theta-e A_\mu)=\xi_\mu =\epsilon_{\mu\nu\rho}\d^\nu a^\rho.
\ee
It leads to a duality relation exchanging the electric current $j_\mu =ev^2\d_\mu\theta$ with the 
vortex current
\be
j^\mu_{\rm vortex}=\frac{e}{2\pi}\epsilon^{\mu\nu\rho}\d_\nu \xi_\rho\;,
\ee
which is the Poincar\'{e} dual of the electric current, 
\be
j^\mu_{\rm vortex}=\frac{1}{2\pi v^2}\epsilon^{\mu\nu\rho}\d_\nu j_\rho\;,
\ee
justifying the name particle-vortex duality. The details of the duality are reviewed in the appendix, but for our purposes 
we will not need more than the above.

\subsection{Maxwell-scalar theory}

We will be interested in general 2+1 dimensional theories with gauge fields, for which we can calculate transport 
coefficients, in particular the conductivity matrix. We therefore start with a Maxwell gauge field, whose kinetic function
$K(\phi)$ is 
defined by a real scalar with canonical kinetic term, 
\begin{equation}
\mathcal{L}_{\phi a_i}=-\frac{1}{2}(\partial_i\phi)^2-V(\phi)-\frac{1}{4}K(\phi)f_{ij}f^{ij}, \phantom{....}i,j=0,1,2,
\end{equation}
where $f_{ij}=\d_i a_j-\d_j a_i$. 

We want to write a dual action corresponding to it so, according to the procedure reviewed in the appendix, in order to 
have a duality valid at the level of the path integral, we write a master action (first order action) that is quadratic in the 
fields. Specifically, we consider $f_{ij}$ now an independent variable and we impose the Bianchi identity $\d_{[i} f_{jk]}=0$, 
that implies $f_{ij}=\d_i a_j-\d_j a_i$, with a Lagrange multiplier $\chi$, thus having 
\be
S_{\rm master}=\int d^{2+1}x\left[-\frac{1}{2}(\partial_i\phi)^2-V(\phi)-\frac{1}{4}K(\phi)f_{ij}f^{ij}
+\frac{1}{2}\epsilon^{ijk}\chi \d_i f_{jk}\right].
\ee
Then indeed, eliminating the Lagrange multiplier $\chi$ through its equation of motion, we come back to the 
original action of ${\cal L}_{\phi a}$. 

If instead we eliminate the field $f_{ij}$ through its simple equation of motion, we obtain the dual action in terms of the 
Lagrange multiplier $\chi$, promoted to a real scalar field. We first partially integrate the term with $\chi$ (ignoring the 
boundary term) to $-\frac{1}{2}\d_i \chi f_{jk}$, then we obtain the equation of motion 
\be
K(\phi) f^{ij}+\epsilon^{ijk}\d_k \chi=0\;,
\ee
and then we replace the resulting 
\be
f^{ij}=-\frac{1}{K(\phi)}\epsilon^{ijk}\d_k \chi
\ee
in the master action, obtaining the dual Lagrangian
\be
{\cal L}^{\rm dual}_{\phi \chi}=-\frac{1}{2}(\d_i \phi)^2-V(\phi)-\frac{1}{2K(\phi)}(\d_i \chi)^2.
\ee

As in the generic particle-vortex duality 
case before, we have traded a gauge field $a_i$ for a real scalar $\chi$. Moreover, as is appropriate for a 
duality, we have inverted the value of the coupling, from $g^2=1/K(\phi)$ for $a_i$ to $g'^2=K(\phi)$ for $\chi$.

\subsection{Including a Chern-Simons term}

Next, we introduce a Chern-Simons term for the Maxwell field, with coefficient $-\Theta/(2\pi)$,
\begin{equation}
\mathcal{L}_{\phi a_i; \Theta}=-\frac{1}{2}(\partial_i\phi)^2-V(\phi)-\frac{1}{4}K(\phi)f_{ij}f^{ij}
-\frac{\Theta}{4 \pi}\epsilon^{ijk}a_i f_{jk}.
\end{equation}

Now however, we cannot continue as before, imposing the Bianchi identity with a Lagrange multiplier $\chi$
for a master action in terms of an independent field $f_{ij}$, since the 
field $a_i$ already appears in the Chern-Simons term, and there is no way to equate it with the gauge field $a'_i$
that solves the Bianchi identity. 

One possibility is to use the duality of this ``topologically massive'' theory \cite{Deser:1984kw} to a ``self-dual in 
odd dimensions'' theory \cite{Townsend:1983xs}, as done in \cite{Murugan:2016zal} to related topological insulators to 
topological superconductors. We start with this, slightly extending the analysis of  \cite{Murugan:2016zal}. 

We first define 
\be
F^i\equiv \epsilon^{ijk}\d_j a_k\;,
\ee
thus rewriting ${\cal L}_{\phi a_i; \Theta}$ as
\be
\mathcal{L}'_{\phi a_i; \Theta}=-\frac{1}{2}(\partial_i\phi)^2-V(\phi)+\frac{K(\phi)}{2}F_i F^i
-\frac{\Theta}{2 \pi}a_i F^i.
\ee

Next, we write a master action for the Maxwell term, replacing it with an independent field $f_i$, as 
\be
S_{\rm master}^\Theta=\int d^{2+1}x\left[-\frac{1}{2}(\partial_i\phi)^2-V(\phi)
-\frac{K(\phi)}{2}f_i f^i +K(\phi) f_i F^i -\frac{\Theta}{2\pi}F^i a_i \right].
\ee
We can check that the $f_i$ equation of motion is $f^i = F^i$, so we are back to the original action. 

If instead we solve the equation of motion of $a_i$, which is 
\be
\epsilon^{ijk}\d_j \left(K(\phi)f_k\right)=\epsilon^{ijk}\d_j\left(\frac{\Theta}{\pi}a_k\right)\;,
\ee
by 
\be
f_i =\frac{\Theta}{\pi K(\phi)}a_i\;,
\ee
and replace in the master action, we obtain the dual Lagrangian,
\be
{\cal L}^{\rm dual}_{\phi a_i;\Theta}=-\frac{1}{2}(\d_i \phi)^2-V(\phi)-\left(\frac{\Theta}{\pi}\right)^2\frac{a_i a^i}{2
K(\phi)}+\frac{\Theta}{2\pi}\epsilon^{ijk}a_i \d_j a_k.
\ee

However, in this case it is less transparent what the action of the duality on the parameters of the theory, $K(\phi)$ and 
$\Theta/\pi$, is, as the dual action is of a different type than the original one. 

But there is another way of doing the duality that is closer to our goal, since it exchanges a gauge field with another 
gauge field, and is closer in spirit to what we will do in 3+1 dimensions. Instead of imposing the Bianchi identity for 
$f_{ij}$, which would leave no control over what is the relation of the gauge field $a'_i$ to the $a_i$ in the Chern-Simons 
term, we impose the constraint that $f_{ij}=\d_i a_j-\d_j a_i$, with a Lagrange multiplier ${\cal A}_i$, to obtain the master
action
\bea
{S'}_{\rm master}^{\Theta}&=&\int d^{2+1}x \left[-\frac{1}{2}(\partial_i\phi)^2-V(\phi)
-\frac{1}{4}K(\phi)f_{ij}f^{ij}\right.\cr
&&\left.-\frac{\Theta}{4 \pi}\epsilon^{ijk}a_i f_{jk} +C \epsilon^{ijk}{\calç A}_i(f_{jk}-\d_j a_k +\d_k a_j)\right]\;,
\eea
where $C$ is a constant, which for the moment is not fixed, but should be. 
Then if we vary with respect to the Lagrange multiplier ${\cal A}_i$, we get back to the original action. 

But if instead we vary with respect to the gauge field $a_i$, we obtain an equation of motion that solves for $f_{ij}$ as
\be
f_{ij}=-\frac{4\pi}{\Theta}C (\d_i {\cal A}_j-\d_j {\cal A}_i).
\ee
When putting it back inside the master action, we obtain the dual Lagrangian,
\be
{\cal L}^{\rm dual}_{\phi {\cal A}_i}=-\frac{1}{2}(\d_i \phi)^2-V(\phi)+\left(\frac{4\pi}{\Theta}\right)^2C^2
\left[-\frac{1}{4}K(\phi)\tilde f_{jk}^2-\frac{\Theta}{4\pi}\epsilon^{ijk}{\cal A}_i \tilde f_{jk}\right]\;,
\ee
where $\tilde f_{ij}=\d_i {\cal A}_j-\d_j {\cal A}_i$. We note then that the dual action has the same form as the original
one, except with modified coefficients,
\bea
K&\rightarrow& K'=\left(\frac{4\pi C}{\Theta}\right)^2K\cr
\frac{\Theta}{\pi}&\rightarrow & \frac{\Theta'}{\pi}=\left(\frac{4\pi C}{\Theta}\right)^2\frac{\Theta}{\pi}.
\eea

We then notice if we choose $C$ to equal to the coefficient of the Chern-Simons term, normalized by the common 
factor of it and the Maxwell term, i.e.,
\be
C=\frac{\frac{\Theta}{4\pi}}{\sqrt{K^2+\frac{\Theta^2}{\pi^2}}}\;,
\ee
then the action on the duality on the coefficients is 
\bea
K&\rightarrow& K'=\frac{K}{K^2+\frac{\Theta^2}{\pi^2}}\cr
\frac{\Theta}{\pi}&\rightarrow & \frac{\Theta'}{\pi}=\frac{\frac{\Theta}{\pi}}{K^2+\frac{\Theta^2}{\pi^2}}\;,
\eea
which will be also what we find from 3+1 dimensions, with $\Theta_{2+1}=4\Theta_{3+1}$.

\subsection{Field theory action as response action and effect of duality on conductivities}

The actions considered before, with a gauge field $a_i$, could a priori be considered actions for a fundamental, 
electromagnetic, gauge field, whose variation would give the equation of motion for $a_i$. But in the case we are interested 
in, of a nonzero Chern-Simons term (which as we saw is qualitatively different from the case without such a term, at 
least as far as the duality is concerned), a better interpretation is as {\em response actions}, encoding the response of the 
material to an {\em external} gauge field $a_i$ (whose dynamics is therefore not encapsulated by the action we consider). 

Indeed, it is well known that the quantum Hall effect is encapsulated in the Chern-Simons action (with a quantized coefficient),
understood as a topological response action \cite{Qi:2008ew}. The Hall current, $j^a=\sigma_H\epsilon^{ab}E_b$, where 
$a,b=1,2$ are spatial indices, also implies (using current conservation and the Maxwell equations)
\be
\d_0 \rho=-\d_a j^a=-\sigma_H \epsilon^{ab} \d_a E_b =\sigma_H \d_0 B \Rightarrow \rho(B)-\rho_0=\sigma_B\;,
\ee
and together, these equations form the topological response action coming from the Chern-Simons action, 
\be
j^i=\sigma_H \epsilon^{ijk}\d_j a_k=\frac{\delta S_{\rm CS}}{\delta a_i}.
\ee

We could in principle include also the Maxwell term in this analysis, thus considering that the Maxwell term also encodes the 
response of the material, this time the longitudinal response, provided we take the point that the 
electric field varies microscopically in time. Indeed, then we would write for the spatial components of the total 
current, obtained from the action viewed as a response action, so 
\be
j^a=\frac{\delta S}{\delta a_a}=K \d_j f^{ja}-\frac{\Theta}{2\pi}\epsilon^{aij}\d_i a_j\simeq 
K \d_0 f^{0a}-\frac{\Theta}{2\pi} \epsilon^{ab}E_b\simeq \frac{K}{\tau}\langle E_a\rangle -\frac{\Theta}{2\pi}\epsilon^{ab}
E_b\;,
\ee
where we have assumed a time variation over a time scale $\tau$. Then $K/\tau$ acts as normal conductivity
$\sigma_{ij,||}=\sigma_{||}\delta_{ij}$ and $\Theta/(2\pi)$ as Hall conductivity $\sigma_{ij,\perp}=\sigma_H\epsilon_{ij}$.

We finally deduce that the action of particle-vortex duality on the normal and Hall conductivities is 
\bea
\sigma_{||}&\rightarrow & \sigma'_{||}=\frac{\sigma_{||}}{\sigma_{||}^2+\sigma_H^2}\cr
\sigma_H& \rightarrow & \sigma'_H=-\frac{\sigma_H}{\sigma_{||}^2+\sigma_H^2}.
\eea
This is the same action as was found by Burgess and Dolan \cite{Burgess:2000kj} and also by Son \cite{Son:2015xqa}, 
which in terms of $\sigma=\sigma_{xy}+i\sigma_{xx}$, where as we saw $\sigma_H=\sigma_{xy}$ and $\sigma_{||}=
\sigma_{xx}$, is 
\be
\tilde \sigma=-\frac{1}{\sigma}.
\ee

\section{S-duality in 3+1 dimensional theories coupled to gravity and effect on conductivities}

The next issue we want to describe is the action of S-duality, or particle-monopole duality, in 3+1 dimensional theories 
for Maxwell vector fields coupled to gravity, and a scalar field defining the kinetic terms.
The reason is that we would like to 
describe the effect of the particle-vortex duality in a gravity dual, and as we argued, particle-vortex duality is mapped to 
S-duality.\footnote{The idea of an S-duality, 
or $Sl(2,\mathbb{Z})$ invariant phenomenological gravity dual has been used a lot, see for instance \cite{Lippert:2014jma}.} 

The metric backgrounds we will consider then must be asymptotically AdS, and we also assume the existence of a black hole
horizon inside the bulk, in order to have temperature and thermodynamic properties, including entropy, as well as 
transport properties like conductivity, which will be our main focus.

Near the black hole horizon, we consider 
\begin{equation}
g_{00}=-\frac{c_0}{r-r_0}, \phantom{...........} g_{rr}=c_r(r-r_0), 
\end{equation}
where $c_0$ and $c_r$ are constants.

\subsection{S-duality}

In this subsection we define the usual S-duality transformation, with the only generalization of a kinetic term for the 
abelian (Maxwell) vector field that depends on a scalar field. Thus, consider the Einstein-Maxwell-dilaton action 
\begin{equation}
S_{\rm g-A-\phi}=\int d^4x\sqrt{-g}\left[\frac{1}{2\kappa_N^2}\left(R-\frac{6}{L^2}\right)
-\frac{1}{2}(\partial_{\mu}\phi)^2-V(\phi)-\frac{K(\phi)}{4}g^{\mu\rho}
g^{\nu\sigma}F_{\mu\nu}F_{\rho\sigma}\right]\;,\label{model3}
\end{equation}
where $F_{\mu\nu}=\d_\mu A_\nu -\d_\nu A_\mu$. 
In most of the following, we put $2\kappa_N^2=1$. Here $L$ is the radius of the AdS background solution, considering that
at the minimum, $V(\phi_{\rm min})=0$. 

As usual, for dualization of the action, we consider $F_{\mu\nu}$ to be an independent field, and 
add the Bianchi identity for it as a constraint with a Lagrange multiplier $B_\mu$, obtaining a master action 
\be
S_{\rm master}=S_{\rm g-A-\phi}+\frac{1}{2}\int d^4x \epsilon^{\mu\nu\rho\sigma}B_\mu \d_\nu F_{\nu\rho}. 
\ee
If we solve for the constraint of the Lagrange multiplier $B_\mu$, we go back to the original action. 
We can instead solve for $F_{\mu\nu}$ and, defining 
\be
\tilde F_{\mu\nu}\equiv \d_\mu B_\nu -\d_\nu B_\mu\;,
\ee
and we obtain the equation of motion 
\be
F^{\mu\nu}=\frac{1}{2\sqrt{-g}K(\phi)}\epsilon^{\mu\nu\rho\sigma}\tilde F_{\rho\sigma}.
\ee
Replacing in the action and using $\epsilon_{\mu\nu\lambda \tau}\epsilon^{\lambda\tau\rho\sigma}=4g\delta_{\mu\nu}
^{\rho\sigma}$, we obtain the dual action 
\be
S_{\rm g-B-\phi}^{\rm dual}=\int d^4x\sqrt{-g}\left[\frac{1}{2\kappa_N^2}\left(R-\frac{6}{L^2}\right)
-\frac{1}{2}(\partial_{\mu}\phi)^2-V(\phi)-\frac{1}{4 K(\phi)}g^{\mu\rho}
g^{\nu\sigma}\tilde F_{\mu\nu}\tilde F_{\rho\sigma}\right]\;,
\ee
so just the original action, inverting the coupling function, $K(\phi)\rightarrow 1/K(\phi)$ and exchanging the field
strength $F_{\mu\nu}$ with the dual field strength $\tilde F_{\mu\nu}$.

\subsection{S-duality with theta term}

Next we introduce a theta term to the 4-dimensional action, obtaining 
\bea
S_{\rm g-A-\phi}^\Theta&=&\int d^4x\left\{\sqrt{-g}\left[\frac{1}{2\kappa_N^2}\left(R-\frac{6}{L^2}\right)
-\frac{1}{2}(\partial_{\mu}\phi)^2-V(\phi)-\frac{K(\phi)}{4}g^{\mu\rho}
g^{\nu\sigma}F_{\mu\nu}F_{\rho\sigma}\right]\right.\cr
&&\left.+\frac{\Theta}{2\pi}\epsilon^{\mu\nu\rho\sigma}F_{\mu\nu}F_{\rho\sigma}\right].
\eea
As before, we write a master action by makng the field $F_{\mu\nu}$ independent, and imposing its Bianchi identity 
with a Lagrange multiplier $B_\mu$, 
\be
S_{\rm master}^\Theta=S_{\rm g-A-\phi}^\Theta
+\frac{1}{2}\int d^4x \epsilon^{\mu\nu\rho\sigma}B_\mu \d_\nu F_{\nu\rho}. 
\ee
Indeed, then if we solve for $B_\mu$ we get back to the original action, or if we solve for $F_{\mu\nu}$, we get the 
equation of motion 
\be
\frac{\sqrt{-g}K(\phi)}{2}F^{\mu\nu}=\frac{\Theta}{\pi}\epsilon^{\mu\nu\rho\sigma} F_{\rho\sigma}+
\frac{1}{2}\epsilon^{\mu\nu\rho\sigma}
\d_\rho B_\sigma. 
\ee

Defining 
\be
C^{\mu\nu\rho\sigma}\equiv \frac{\sqrt{-g}K(\phi)}{2}g^{[\mu|\rho}g^{\nu]\sigma}-\frac{\Theta}{\pi}
\epsilon^{\mu\nu\rho\sigma}\;,
\ee
we solve for $F_{\mu\nu}$ in terms of the dual $\tilde F_{\mu\nu}=\d_\mu B_\nu-\d_\nu B_\mu$ as 
\be
F_{\mu\nu}=\tilde C_{\mu\nu\rho\sigma}\frac{1}{4}\epsilon^{\rho\sigma\a\b}\tilde F_{\a\b}\;,
\ee
where $\tilde C_{\mu\nu\rho\sigma}=
(C^{\mu\nu\rho\sigma})^{-1}$ is the inverse matrix.

We proceed by parametrizing the inverse matrix as 
\be
\tilde C_{\mu\nu\rho\sigma}=\frac{a}{\sqrt{-g}}g_{[\mu|\rho}g_{\nu]\sigma}+\frac{b}{(-g)}\epsilon_{\mu\nu\rho\sigma}\;,
\ee
so that
\be
F_{\mu\nu}=\frac{a}{\sqrt{-g}}\frac{1}{4}{\epsilon_{\mu\nu}}^{\rho\sigma}\tilde F_{\rho\sigma}
-b \tilde F_{\mu\nu}\;,\label{dualityrel}
\ee
where as usual $\epsilon^{0123}=+1$ and $\epsilon_{\mu\nu\rho\sigma}$ has indices lowered with the metric.
Then we impose ($\delta^{\mu\nu}_{\rho\sigma}\equiv \frac{1}{2}(\delta^\mu_\rho\delta^\nu_\sigma-\delta^\mu_\sigma
\delta^\nu_\rho)$)
\be
C^{\mu\nu\a\b}\tilde C_{\a\b \rho\sigma}=\delta^{\mu\nu}_{\rho\sigma}\;,
\ee
which gives 
\be
\left(\frac{K(\phi)a}{2}+\frac{4\Theta b}{\pi}\right)\delta^{\mu\nu}_{\rho\sigma}+\left(-\frac{\Theta a}{\pi \sqrt{-g}}
+\frac{K(\phi)b}{2\sqrt{-g}}\right){\epsilon^{\mu\nu}}_{\rho\sigma}=\delta^{\mu\nu}_{\rho\sigma}\;,
\ee
with the solution 
\be
a=\frac{K(\phi)/2}{(K(\phi)/2)^2+(2\Theta/\pi)^2};\;\;\;
b=\frac{\Theta/\pi}{(K(\phi)/2)^2+(2\Theta/\pi)^2}.
\ee
Then the master action, written as 
\bea
S_{\rm master}^\Theta
&=&\int d^4x\left\{\sqrt{-g} \left[\frac{1}{2\kappa_N^2}\left(R-\frac{6}{L^2}\right)
-\frac{1}{2}(\partial_{\mu}\phi)^2-V(\phi)\right]\right.\cr
&&\left.-\frac{1}{2}C^{\mu\nu\rho\sigma}F_{\mu\nu}F_{\rho\sigma}+\frac{1}{4}
\epsilon^{\mu\nu\rho\sigma}F_{\mu\nu}
\tilde F_{\rho\sigma}\right\}
\eea
becomes the dual action 
\bea
S^{\rm dual,\Theta}_{\rm g-B-\phi}&=&\int d^4x \left\{\sqrt{-g} \left[\frac{1}{2\kappa_N^2}\left(R-\frac{6}{L^2}\right)
-\frac{1}{2}(\partial_{\mu}\phi)^2-V(\phi)\right]\right.\cr
&&\left.+\frac{1}{32}\left({\epsilon_{\mu\nu}}^{\rho\sigma}{\tilde C_{\rho\sigma}}^{\lambda\tau}{\epsilon_{\lambda
\tau}}^{\eta\theta}\right)\tilde F^{\mu\nu}\tilde F_{\eta\theta}\right]\cr
&=&\int d^4x \left\{\sqrt{-g} \left[\frac{1}{2\kappa_N^2}\left(R-\frac{6}{L^2}\right)
-\frac{1}{2}(\partial_{\mu}\phi)^2-V(\phi)\right]\right.\cr
&&\left.+\left[-\sqrt{-g}\frac{a}{8}\tilde F_{\mu\nu}\tilde F^{\mu\nu}
-\frac{b}{8}\epsilon^{\mu\nu\rho\sigma}\tilde F_{\mu\nu}\tilde F_{\rho\sigma}\right]\right\}\cr
&=&\int d^4x \left\{\sqrt{-g} \left[\frac{1}{2\kappa_N^2}\left(R-\frac{6}{L^2}\right)
-\frac{1}{2}(\partial_{\mu}\phi)^2-V(\phi)\right]\right.\cr
&&\left.+ \frac{1}{K(\phi)^2+(4\Theta/\pi)^2}\left[-\sqrt{-g}\frac{K(\phi)}{4}\tilde F_{\mu\nu}\tilde F^{\mu\nu}
-\frac{\Theta}{2\pi}\epsilon^{\mu\nu\rho\sigma}\tilde F_{\mu\nu}\tilde F_{\rho\sigma}\right]\right\}.
\eea
This is the same as the original action, just with the replacement
\bea
K&\rightarrow& K'=\frac{K}{K^2+\frac{(4\Theta)^2}{\pi^2}}\cr
\frac{4\Theta}{\pi}&\rightarrow & \frac{4\Theta'}{\pi}=-\frac{\frac{4\Theta}{\pi}}{K^2+\frac{(4\Theta)^2}{\pi^2}}\;,
\eea
the same as we obtained in the 2+1 dimensional calculation (except for a rescaling of $K$ and $\Theta$). Note however 
that the $K,\Theta$ parameters are now defined in 3+1 dimensions, therefore {\em a priori} differently.

\subsection{Effect on 2+1 dimensional conductivities from asymptotically AdS black holes}

We next consider the set-up for an asymptotically AdS black hole solution of the 3+1 dimensional action, and 
the usual calculation of conductivities following the membrane paradigm in \cite{Iqbal:2008}, influenced by the earlier KSS 
work  \cite{Kovtun:2003wp}. The usual boundary term at the horizon of the black hole (considering a radial
foliation, with horizon at $r=r_0$) is defined to be equal to $\int d^{2+1}x j^i A_i$ ($i=0,1,2$), 
so that the current is defined as 
\be
j^i=\frac{\delta S_{3+1}}{\delta \d_r A_i}\;,
\ee
and on the original action with a theta term, we obtain 
\be
j^i=-K(\phi)\sqrt{-g}F^{ri}+\frac{4\Theta}{\pi}\epsilon^{ijk}F_{jk}.
\ee
Here $g$ is the determinant of the 3+1 dimensional metric, written as $g=\gamma g_{rr}$, with $\gamma$ the determinant
of the 2+1 dimensional metric. Moreover, as usual, the condition of regularity at the horizon amounts to the use of the 
Eddington-Finkelstein variable
\be
dv=dt+\sqrt{-\frac{g_{rr}}{g_{00}}}dr\;,
\ee
implying 
\be
\partial_rA_i=\sqrt{-\frac{g_{rr}}{g_{00}}}\partial_0A_i, \phantom{........} r \rightarrow r_0,
\ee
or, with the radial gauge choice $A_r=0$, 
\be
F_{ri}=\sqrt{-\frac{g_{rr}}{g_{00}}}F_{0i}.\label{frf0i}
\ee
This relation will also be derived more rigorously from the Maxwell equations, in a more general context, in section 5.1. 
This allows us to trade $F_{ri}$ for the electric field $F_{0i}$, and obtain the current
\be
j^i=K(\phi)\frac{\sqrt{-g}}{\sqrt{-g_{rr}g_{tt}}}g^{ii}F_{it}+\frac{4\Theta}{\pi}\epsilon^{ijk}F_{jk}\;,
\ee
with no sum over $i=x,y$. Since (for a diagonal metric near the horizon, which we assume) $\sqrt{g}=\sqrt{g_{rr}g_{tt}}
g_{ii}$ (with no sum over $i$), we finally obtain 
\bea
j^x&=& -K(\phi)F_{0x}-\frac{4\Theta}{\pi}F_{0y}\cr
j^y&=& -K(\phi)F_{0y}+\frac{4\Theta}{\pi}F_{0x}.\label{currents3}
\eea
Since $F_{0x}=-E_x$ and $F_{0y}=-E_y$, and in general the conductivity is a matrix defined by $j^a=\sigma^{ab}E_b$, 
for $a,b=x,y$, we obtain 
\be
\sigma_{xx}=K(\phi)\;,\;\;\; \sigma_{xy}=\frac{4\Theta}{\pi}.
\ee

Then after the duality transformation, we find that the new conductivities satisfy
\begin{eqnarray}
\sigma^{\prime}_{xx}&=& \frac{\sigma_{xx}}{\sigma^{2}_{xx}+\sigma^{2}_{xy}},\\
\sigma^{\prime}_{xy}&=& -\frac{\sigma_{xy}}{\sigma^{2}_{xx}+\sigma^{2}_{xy}}.
\end{eqnarray}
Then forming as before $\sigma=\sigma_{xy}+i\sigma_{xx}$, we find the same S-duality transformation as 
calculated in 2+1 dimensions, 
\begin{equation}
\sigma^{\prime} = -\frac{1}{\sigma}.
\end{equation}
Unlike the action of the duality on $K$ and $\Theta$, which were defined {\em a priori} differently in 2+1 and in 3+1 
dimensions (and with different normalizations), 
the action on the physical conductivities is unambiguous, and we see that indeed we obtain the same result, as we expected 
due to the existence of the AdS/CFT correspondence.

\section{Relating the 3+1 dimensional case to the 2+1 dimensional case through AdS/CFT}

We now show that indeed, AdS/CFT should relate the two pictures, as it obviously does.
It was shown in  \cite{Murugan:2014sfa}, in the case that the 3+1 dimensional action had only a simple Maxwell term 
(and the 2+1 dimensional action had no Chern-Simons term), that the Maxwell duality (S-duality) relation in 3+1 dimensional
AdS space reduces to a set of relations that imply both the particle-vortex duality relation, and the equation of motion for 
the Maxwell field, defining the evolution in the radial direction. Here we generalize this result to the case of the theta term
and of the asymptotically AdS black hole, and also we show that a similar analysis can be performed near the horizon 
of the black hole. 

\subsection{Analysis at the boundary}

In principle we should consider the expansion of the metric near the boundary, as being AdS space plus corrections. 
However, since the analysis is more complicated, but morally nothing changes, we will instead pretend that we 
have just AdS space, with no corrections, everywhere, with metric
\be
ds^2=\frac{-dt^2+dx^2+dy^2+dr^2}{r^2}.
\ee
We expand around the boundary of this AdS space, at $r=0$.
Consider the duality relation (\ref{dualityrel}) and use the radial gauge $A_r=0$, which becomes explicitly (substituting the 
metric of AdS and using $\epsilon^{0rab}=\epsilon^{ab}$, and redefining $a/2\rightarrow a$)
\bea
F_{0a}&=&a\d_r \epsilon^{ab}\tilde A_b-b\tilde F_{0a}\cr
\d_r A_a&=&a\epsilon^{ab}\tilde F_{0b}-b \d_r \tilde A_a.\label{dual1}
\eea
We also write the inverse duality relation,
\be
\tilde F_{\mu\nu}=\frac{K(\phi)}{2\sqrt{-g}}{\epsilon_{\mu\nu}}^{\rho\sigma}F_{\rho\sigma}
-\frac{4\Theta}{\pi}F_{\mu\nu}\;,
\ee
which becomes
\bea
\tilde F_{0a}&=& K(\phi) \epsilon^{ab}\d_r A_b -\frac{4\Theta}{\pi}F_{0a}\cr
\d_r\tilde A_a&=&K(\phi)\epsilon^{ab}F_{0b}-\frac{4\Theta}{\pi}\d_r A_a.\label{dual2}
\eea
We can also similarly express $F_{ab}, A_0$ and $\tilde F_{ab}, \tilde A_0$, to generalize the relations to 
$F_{ij},A_i$ and $\tilde F_{ij},\tilde A_i$.

Expanding in $r$, 
\bea
A_i&=&\sum_{n\geq 0}\frac{r^n}{n!} a_i^{(n)}\cr
F_{ij}&=&\sum_{n\geq 0}\frac{r^n}{n!}f_{ij}^{(n)}\cr
\tilde F_{ij}&=&\sum_{n\geq 0}\frac{r^n}{n!}\tilde f_{ij}^{(n)}\;,
\eea
and substituting in (\ref{dual1}) and (\ref{dual2}) and their generalizations, we obtain (using $\epsilon^{ab}=\epsilon^{0ab}
=-\epsilon^{a0b}$)
\bea
f_{ij}^{(n)}&=&-a\epsilon_{ijk}\tilde a_k^{(n+1)}-b\tilde f_{ij}^{(n)}\label{f}\\
a_i^{(n+1)}&=&-b\tilde a_i^{(n+1)}-\frac{a}{2}\epsilon_{ijk}\tilde f_{jk}^{(n)}\label{a}\\
\tilde f_{ij}^{(n)}&=&-K(\phi)\epsilon^{ijk}a_k^{(n+1)}-\frac{4\Theta}{\pi}f_{ij}^{(n)}\label{tildef}\\
\tilde a_i^{(n+1)}&=&-\frac{4\Theta}{\pi}a_i^{(n+1)}-\frac{K(\phi)}{2}\epsilon^{ijk}f_{jk}^{(n)}\label{tildea}.
\eea

Writing $f_{ij}^{(n+1)}=\d_i a_j^{(n+1)}-\d_j a^{(n+1)}$ from both (\ref{f}) and from (\ref{a}), using 
$2\d_{[i}\epsilon_{j]kl}=2\d_{[l}\epsilon_{k]ij}$, and equating the 
two results, we obtain 
\be
\frac{1}{2}\d^j \tilde f_{ji}^{(n)}=-\tilde a_i ^{(n+2)}.
\ee
Similarly, writing $\tilde f_{ij}^{(n+1)}=\d_i \tilde a_j^{(n+1)}-\d_j \tilde a_i^{(n+1)}$ from both (\ref{tildef}) and 
from (\ref{tildea}) and equating the two results, we obtain 
\be
\frac{1}{2}\d^j f_{ji}^{(n)}=-a_i ^{(n+2)}.
\ee
These two relations are the same ones obtained in \cite{Murugan:2014sfa}, so the presence of the theta term doesn't affect 
them. It means, as stated in \cite{Murugan:2014sfa}, that we can freely give either $a_i^{(0)}$ and $\tilde a_i^{(0)}$, 
or $a_i^{(0)}$ and $a_i^{(1)}$, and then the higher orders are found from the duality relation, standing in for the 
equations of motion. The duality relation for $a_i^{(1)}$ in terms of $\tilde a_i^{(1)}$ and $\tilde a_i^{(0)}$, 
\be
a_i^{(1)}=-b\tilde a_i^{(1)}-a \epsilon_{ijk}\d_j \tilde a_k^{(0)}\;,
\ee
as well as the reverse one, 
\be
\tilde a_i^{(1)}=-\frac{4\Theta}{\pi}a_i ^{(1)}-K(\phi)\epsilon_{ijk}\d_j a_k^{(0)}\;,
\ee
are seen to be the same ones from the 2+1 dimensional case, as advertised.

Note that half of the equations in (\ref{f},\ref{a},\ref{tildef},\ref{tildea}) 
are redundant, substituting ones in others we obtain consistency 
conditions that are satisfied.

\subsection{Analysis at the horizon}

The analysis at the horizon is somewhat similar, just that we now have 
\be
g_{00}=-\frac{c_0}{r-r_0}\;,\;\;\; g_{rr}=c_r(r-r_0)\;,
\ee
which means that the duality relation (\ref{dualityrel}) becomes 
\bea
F_{0a}&=& a\frac{\a}{r-r_0}\epsilon_{ab}\d_r\tilde A_b-b\tilde F_{0a}\cr
\d_r A_a&=&\frac{a(r-r_0)}{\a}\epsilon_{ab}\tilde F_{0b}-b\d_r \tilde A_a\;,
\eea
where we have defined $\a\equiv \sqrt{\frac{c_0}{c_r}}$, and similar relations for $F_{ab}$ and $\d_r A_0$. 
Similarly, the inverse duality relation becomes 
\bea
\tilde F_{0a}&=& \frac{\a K(\phi)}{r-r_0}\epsilon_{ab} \d_r A_b-\frac{4\Theta}{\pi}F_{0a}\cr
\d_r \tilde A_a&=&\frac{K(\phi)(r-r_0)}{\a}\epsilon_{ab}\tilde F_{0b}-\frac{4\Theta}{\pi}\d_r  A_a\;,
\eea
and similar relations for $\tilde F_{ab}$ and $\d_r \tilde A_0$. 
We now define an expansion in terms of $\tilde r=r-r_0$, 
\bea
A_i&=&\sum_{n\geq 0}\frac{\tilde r^n}{n!} a_i^{(n)}\cr
F_{ij}&=&\sum_{n\geq 0}\frac{\tilde r^n}{n!}f_{ij}^{(n)}\cr
\tilde F_{ij}&=&\sum_{n\geq 0}\frac{\tilde r^n}{n!}\tilde f_{ij}^{(n)}\;,
\eea
just that, unlike the boundary, the duality relations have now extra factors of $\tilde r$. 

Then by substituting the expansions in the generalizations of the duality relations above, we obtain 
\bea
f_{ij}^{(n)}&=&-b \tilde f_{ij}^{(n)}-a\a \epsilon_{ijk}\frac{\tilde a_k^{(n+2)}}{n+1}\label{fh}\\
\frac{a_i^{(n+2)}}{n+1}&=&-b \frac{\tilde a_i^{(n+2)}}{n+1}-\frac{a}{\a}\frac{1}{2}\epsilon_{ijk}\tilde f_{jk}^{(n)}
\label{ah}\\
\tilde f_{ij}^{(n)}&=&-\frac{4\Theta}{\pi}f_{ij}^{(n)}-\frac{K(\phi)\a}{n+1}\epsilon_{ijk}a_k^{(n+2)}\label{tildefh}\\
\frac{\tilde a_i^{(n+2)}}{n+1}&=& -\frac{4\Theta}{\pi}\frac{a_i^{(n+2)}}{n+1}-\frac{K(\phi)}{\a}\frac{1}{2}
\epsilon_{ijk}f_{jk}^{n)}\label{tildeah}.
\eea

Writing $f_{ij}^{(n+2)}=\d_i a_j^{(n+2)}-\d_j a_i^{(n+2)}$ from both (\ref{ah}) and from (\ref{fh}), and equating 
the two results, we obtain 
\be
\frac{1}{2}\d^k \tilde f_{kl}^{(n)}=-\frac{\a^2}{(n+1)(n+3)}\tilde a_l ^{(n+4)}.
\ee
Similarly, writing $\tilde f_{ij}^{(n+2)}=\d_i \tilde a_j^{(n+2)}-\d_j a_i^{(n+2)}$ from both (\ref{tildeah}) and 
from (\ref{tildefh}) and equating the two results, we obtain 
\be
\frac{1}{2}\d^k f_{kl}^{(n)}=-\frac{\a^2}{(n+1)(n+3)}a_l^{(n+4)}.
\ee

Thus the only new feature is that now we relate $a_l^{(n)}$ with $a_l^{(n+4)}$, instead of $a_l^{(n+2)}$. 
That means that, {\em by the duality relation alone}, we can freely give $a_i^{(0)}, a_i^{(1)}, a_i^{(2)}$ and 
$a_i^{(3)}$ at the horizon (of course, the equations of motion can further constrain this). 

At the horizon, the duality relations between the freely specified gauge fields contain extra factors of $\a$, 
\bea
a_i^{(2)}&=& -b\tilde a_i^{(2)}-\frac{a}{\a}\epsilon_{ijk}\d_j \tilde a_k^{(0)}\cr
\tilde a_i^{(2)}&=& -\frac{4\Theta}{\pi} a_i^{(2)}-\frac{K(\phi)}{\a}\epsilon_{ijk}\d_j a_k^{(0)}\;,
\eea
but otherwise are the same.

\section{General 3+1 dimensional Einstein-gauge-dilaton action involving quantum corrections}

In this section we will consider the effect of quantum corrections to the 3+1 dimensional gravitational plus vector plus 
scalar action considered in the previous sections. This will introduce extra terms, both in the gravitational action, and 
in the scalar kinetic action. 

As before, we will be interested in metrics that go asymptotically (near the boundary) to AdS space, and have black hole 
horizons in a radial foliation, at some value $r=r_0$. Generically then, we consider a diagonal metric that depends only on the 
radial coordinate, i.e., 
\be
ds^2=g_{00}(r) dt^2+g_{rr}(r) dr^2+\sum_{a=1,2} g_{aa}(r) (dx^a)^2.
\ee

Moreover, we are interested in a more general kinetic term for the Maxwell field
(see \cite{Brigante:2007nu,Myers:2010pk}), of the form (note that even the $\sqrt{-g}$ is considered part of the 
kinetic function)
\be
S_M=\int d^{3+1}x \left[-\frac{1}{4}F_{\mu\nu}X^{\mu\nu\rho\sigma}F_{\rho\sigma}\right]\;,\label{SM}
\ee
where $X^{\mu\nu\rho\sigma}$ is antisymmetric in $(\mu\nu)$ and $(\rho\sigma)$.

Moreover, we have in mind an application to a case similar to the ones from previous sections, like
\be
S_{\rm gauge}=\int d^{3+1}x\left[-\frac{K(\phi(r))}{4}\sqrt{-g} F_{\mu\nu}F^{\mu\nu}+\frac{2\Theta}{\pi}
F_{\mu\nu}\tilde F^{\mu\nu}\right]\;,\label{sgauge}
\ee
so that we can split the kinetic matrix $X$ into a diagonal, and an off-diagonal part, 
\be
X^{\mu\nu\rho\sigma}=AX^{\mu\nu\rho\sigma}\delta_{\mu\nu}^{\rho\sigma}+B \epsilon^{\mu\nu\rho\sigma}\;,
\ee
(no sum over $\mu\nu\rho\sigma$)
where $A$ and $B$ are defined such that they are constants at the horizon $r=r_0$, whereas $X^{\mu\nu\rho\sigma}$ 
doesn't need to be.

\subsection{Formalism for evolution of the abelian field strength in gravitational background}

The equation of motion for the Maxwell field coming from (\ref{SM}) is 
\be
\d_\mu (X^{\mu\nu\rho\sigma}F_{\rho\sigma})=0\label{Max}
\ee
This needs to be supplemented with the Bianchi identity
\be
\d_{[\mu}F_{\nu\rho]}=0\;,
\ee
which implies that $F_{\mu\nu}=\d_\mu A_\nu-\d_\nu A_\mu$.

We also consider gauge field perturbations of definite momenta $q_1,q_2$ in the two spatial boundary direction, so 
\be
A_\mu =A_\mu(t,r)e^{iq_a x^a}.
\ee

Then, considering that the metric itself depends only on $r$ (so also does $X^{\mu\nu\rho\sigma}$), but 
the gauge field perturbation depends on everything, 
the $\nu=t,r,x,y$ components of (\ref{Max}) 
become
\bea
&&\partial_r(AX^{rtrt}F_{rt})-\partial_r(BF_{xy})-iq_1AX^{xtxt}F_{0x}-iq_2AX^{ytyt}F_{0y}+iB(q_1F_{ry}
-q_2F_{rx})=0\cr
&&AX^{trtr}\d_t F_{0r}+B\partial_tF_{xy}+A(iq_1X^{xrxr}F_{xr}+iq_2X^{yryr}F_{yr})-B(iq_1F_{0y}
-iq_2F_{0x})=0\cr
&&\partial_r(AX^{rxrx}F_{rx})+\partial_r(BF_{0y})+AX^{txtx}\partial_tF_{0x}-B\partial_tF_{ry}-iq_2AX^{xyxy}F_{xy}
-iq_2BF_{0r}=0\cr
&&\partial_r(AX^{ryry}F_{ry})-\partial_r(BF_{0x})+AX^{trtr}\partial_tF_{0y}+B\partial_tF_{rx}+iq_1AX^{xyxy}F_{xy}
+iq_1BF_{0r}=0\;,\cr
&&\label{eom}
\eea
and the components of the Bianchi identity {\em without} $y,x,t,r$ become
\bea
&&-\partial_rF_{0x}+\partial_xF_{0r}+\partial_tF_{rx}=0\cr
&&-\partial_rF_{0y}+\partial_yF_{0r}+\partial_tF_{ry}=0\cr
&&\partial_rF_{xy}-\partial_xF_{ry}+\partial_yF_{rx}=0\cr
&&-\partial_xF_{0y}+\partial_yF_{0x}+\partial_tF_{xy}=0.
\eea

Defining lightcone coordinates $x^\pm$ by 
\be
A_{\pm}\equiv \frac{1}{2}(A_x\pm A_y)\;,\;\;\;
\partial_{\pm}\equiv \frac{1}{2}(\partial_x\pm\partial_y).
\ee
and field strength components
\be
F_{xy}=2F_{-+}\;,\;\;\;
F_{0\pm}=\frac{1}{2}\left(F_{0x}\pm F_{0y}\right)\;,\;\;\;
F_{r\pm}=\frac{1}{2}\left(F_{rx}\pm F_{ry}\right)\;,
\ee
and taking $q_1=q_2=q$, so that $\d_-[...]=0$, the Bianchi identities become
\bea
&&\partial_tF_{r-}=\partial_rF_{0-}\label{Bianchi1}\\
&&\partial_tF_{r+}=\partial_rF_{0+}-iqF_{0r}.\label{Br+}\\
&&\partial_tF_{-+}=-iqF_{0-}\label{Btxy}\\
&&\partial_rF_{-+}=-iqF_{r-}\label{Brxy}.
\eea

Taking the time derivative of the difference between the $x$ and $y$ components of the equation of motion in (\ref{eom}), 
and assuming $X^{txtx}=X^{tyty}$ and $X^{rxrx}=X^{ryry}$,
we obtain 
\bea
&&\d_r(AX^{rxrx}\d_tF_{r-})+AX^{txtx}\d_t^2 F_{0-}+\d_r(B\d_t F_{0+})-B \d_t^2F_{r+}\cr
&&-iq(AX^{xyxy}2\d_t F_{-+}+ B \d_t F_{0r})=0.
\eea

Using the Bianchi identities to replace $\d_t F_{r-}=\d_r F_{0-}$ and $\d_t F_{r+}=\d_r F_{0+}-iq F_{0r}$, 
and then taking the $q\rightarrow 0$ limit, we obtain 
\be
\partial^2_tF_{0-}
+\frac{1}{AX^{txtx}}\partial_r(AX^{rxrx}\partial_rF_{0-})+\frac{\d_r B}{AX^{txtx}}\d_t F_{0+}=0\;,
\ee
where the last term can be assumed to be small (by choosing $B$ to vary little near the horizon), giving
\be
\partial^2_tF_{0-}\simeq -\frac{1}{AX^{txtx}}\partial_r(AX^{rxrx}\partial_rF_{0-}).
\ee
Assuming also that 
\be
A^2X^{rxrx}X^{txtx}\equiv -\tilde \Delta
\ee
is finite (so slowly varying) 
and positive at the horizon (in the case of just the Maxwell term with $K(\phi)$ kinetic function, we obtain 
$\tilde \Delta=K^2(\phi)$), we can rewrite it as 
\be
\partial^2_tF_{0-}\simeq \tilde \Delta \left(\frac{1}{AX^{txtx}}\d_r\right)^2F_{0-}\label{intermed}
\ee

 Assuming a small variation in time, and that $\d_t F_{0-} =\Gamma F_{0-}$ and $\left(\frac{1}{AX^{txtx}}\d_r\right)
F_{0-}=a F_{0-}$, we can solve the quadratic equation and afterwards reform the derivatives.

Taking instead the time derivative of the sum of the $x$ and $y$ components of the equation of motion in (\ref{eom}), 
we obtain
\be
\d_r(AX^{rxrx}\d_tF_{r+})+AX^{txtx}\d_t^2 F_{0+}-\d_r(B\d_t F_{0-})+B \d_t^2F_{r-}=0.
\ee
Using the Bianchi identities as before, and taking the limit $q\rightarrow 0$, and also assuming the term with $\d_r B$ 
to be negligible, we obtain 
\be
\d_t^2 F_{0+}\simeq - \frac{1}{AX^{txtx}}\d_r (A X^{rxrx}\d_r F_{0+})\simeq +\tilde\Delta \left(\frac{1}{AX^{txtx}}\d_r
\right)^2 F_{0+}.
\ee
Assuming as before  $\d_t F_{0-} =\Gamma F_{0-}$ and $\left(\frac{1}{AX^{txtx}}\d_r\right)
F_{0-}=a F_{0-}$, taking the square root algebraically, and then reforming the derivatives, we obtain 
\be
\d_t F_{0+}=\sqrt{\tilde \Delta}\frac{1}{AX^{txtx}}\d_r F_{0+}
=\sqrt{-\frac{X^{rxrx}}{X^{txtx}}}\d_t F_{r+}\;,\label{f+}
\ee
where in the last equality we used the Bianchi identity $\d_r F_{0+}\simeq \d_t F_{r+}$. 

We can do the same trick of taking the square root algebraically and then reforming the derivatives in  (\ref{intermed}),
and obtain similarly
\bea
\d_t F_{0-}
&\simeq & \sqrt{-\frac{X^{rxrx}}{X^{txtx}}}\d_t F_{r-}\;,\label{f-}
\eea
where we have used the Bianchi identity  $\d_r F_{0-}\simeq 
\d_t F_{r-}$.

Now taking the sums and differences of (\ref{f+}) and (\ref{f-}), and using $F_{0x}=F_{0+}+F_{0-}$,
$F_{0y}=F_{0+}-F_{0-}$, $F_{rx}=F_{r+}+F_{r-}$ and $F_{ry}=F_{r+}-F_{r-}$, we obtain finally 
\bea
F_{0x}&\simeq & \sqrt{-\frac{X^{rxrx}}{X^{txtx}}}F_{rx}\cr
F_{0y}&\simeq & \sqrt{-\frac{X^{rxrx}}{X^{txtx}}}F_{ry}.\label{frf0}
\eea

For the standard Maxwell kinetic term with prefactor $K(\phi)$, this becomes just
\be
F_{0a}\simeq \sqrt{-\frac{g^{rr}}{g^{tt}}}F_{ra}\;,
\ee
the same formula that was derived from the fact that the nonsingular coordinate at the horizon was the Eddington-Finkelstein 
one, in (\ref{frf0i}).

In order to apply to our case (\ref{sgauge}), we take $AX^{rxrx}=\sqrt{-g}K(\phi)g^{rr}g^{xx}$, $AX^{txtx}=
\sqrt{-g}K(\phi)g^{tt}g^{xx}$, $B=\frac{2\Theta}{\pi}$. Then we get
\bea
F_{0x}&=& \sqrt{-\frac{g^{rr}}{g^{tt}}}F_{rx}\cr
F_{0y}&=& \sqrt{-\frac{g^{rr}}{g^{tt}}}F_{ry}.
\eea

Inverting these relations, we obtain 
\bea
F_{rx}&=& \sqrt{-\frac{g_{rr}}{g_{tt}}}F_{0x}\cr
F_{ry}&=& \sqrt{-\frac{g_{rr}}{g_{tt}}}F_{0y}.\label{invert}
\eea

We next define, in a covariant formalism, the current at the horizon by the variation of the action with respect to 
$F_{\mu\nu}n^\mu$, where $n_\mu$ is a unit vector in the radial direction, so 
\be
j^\nu =\left.n_\mu X^{\mu\nu\rho\sigma}F_{\rho\sigma}\right|_{r=r_0}.
\ee

Then for the spatial components we get
\be
j^a =n_r X^{ra\rho\sigma}F_{\rho\sigma}=n_r\left[X^{rara}F_{ra}+X^{ra0b}F_{0b}\right]\;,
\ee
without sum over $a$ (only over $b$),
which in our case in (\ref{sgauge}) becomes
\be
j^a=K\sqrt{-\frac{g_{tt}}{g_{rr}}}F_{ra}+\frac{4\Theta}{\pi}\epsilon^{ab}F_{0b}.
\ee

Substituting (\ref{invert}) here, we obtain 
\bea
-j^x&=& K(\phi)F_{0x}+\frac{4\Theta}{\pi}F_{0y}\cr
-j^y&=& K(\phi)F_{0y}-\frac{4\Theta}{\pi}F_{0x}\;,
\eea
the same result as (\ref{currents3}) from section 3.3, which implies as before
\be
\sigma_{xx}=K(\phi)\;,\;\;\;
\sigma_{xy}=\frac{4\Theta}{\pi}.
\ee

\subsection{Effect of induced Gauss-Bonnet-Maxwell terms and S-duality}

Using the formalism of \cite{Brigante:2007nu} (itself extending the one in \cite{Kovtun:2003wp}), a dimensional 
reduction of the 4+1 dimensional action with Gauss-Bonnet quantum gravity corrections,
\be
S_{5dGB}=\frac{1}{16\pi G_N}\int d^5x \sqrt{-g}\left[R-2\Lambda+\frac{\lambda_{GB}}{2}L^2
\left(R^2-4R_{MN}R^{MN}+R_{MNPQ}R^{MNPQ}\right)\right]\;,
\ee
along a direction $y$, under the KK ansatz
\be
ds^2=\tilde{g}_{\mu\nu}dx^{\mu}dx^{\nu}+\epsilon^{2\rho}(dy+A_{\mu}dx^{\mu})^2\;,
\ee
that defines the Maxwell field as coming from the off-diagonal metric,
leads to a 4 dimensional quadratic action for the vector potential of the type
\bea
S_{\rm GB}^{\rm vector}&=&\int d^4x\sqrt{-\tilde{g}}\epsilon^{3\rho}\left[-\frac{1}{4}ZF^2-\frac{\lambda_{GB}}{2}
L^2\left\{Y[R^B]^{\mu\nu\rho\sigma}F_{\mu\nu}F_{\rho\sigma}\right.\right.\cr
&&\left.\left.+4([R^B]^{\tilde{y}\mu\tilde{y}\nu}-[R^B]^{\mu\nu})F_{\mu\rho}{F_\nu}^\rho+([R^B]
-2[R^B]^{\tilde{y}\tilde{y}})F^2\right\}\right]\;,
\label{SGB1}
\eea
where tilde refers to the $\tilde g_{\mu\nu}$ metric, and $R^B$ refers to background curvature, i.e., the 5-dimensional 
quantity with $A_\mu=0$. Moreover, for the reduction we obtain $Z=Y=1$, but if we consider more general quantum 
corrections, we can generalize the action with some arbitrary functions of the scalars $Z,Y$.

The action can be put into the general form (\ref{SM}), with $X^{\mu\nu\rho\sigma}=\sqrt{-\tilde g}
e^{3\rho}\hat X^{\mu\nu\rho\sigma}$
and
\bea
\hat{X}^{\mu\nu\rho\sigma}&=&Zg^{[\nu|\sigma}g^{\mu]\rho}+\lambda_{GB}L^2Y\left\{|R^B|^{\mu\nu\rho\sigma}
+4(|R^B|^{\tilde{y}\mu\tilde{y}[\rho}-|R^B|^{\mu[\rho})g^{\nu|\sigma]}\right.\cr
&&\left.+(|R^B|-2|R^B|^{\tilde{y}\tilde{y}})g^{[\nu|\sigma}g^{\mu]\rho}\right\}\cr
&\equiv & Zg^{[\nu|\sigma}g^{\mu]\rho}+\lambda_{GB}L^2YE(g)^{\mu\nu\rho\sigma}  .\label{Xhat}
\eea

We further generalize (\ref{SGB1}) by adding a theta term, and then writing a master action, by turning the $F_{\mu\nu}$
into an independent field, and imposing the Bianchi identity with a Lagrange multiplier $a_\mu$, after which we 
partially integrate, to obtain 
\be
S_{\rm master}^\Theta=\int d^4x \left[-\frac{1}{4}F_{\mu\nu}X^{\mu\nu\rho\sigma}F_{\rho\sigma}
+\frac{\Theta}{2\pi}\epsilon^{\mu\nu\rho\sigma}F_{\mu\nu}F_{\rho\sigma}
-\frac{1}{2}\epsilon^{\mu\nu\rho\sigma}f_{\mu\nu}F_{\rho\sigma}\right]\;,
\ee
where $f_{\mu\nu}=\d_\mu a_\nu-\d_\nu a_\mu$. We can reabsorb the theta term, as before, in a redefinition of the 
$X^{\mu\nu\rho\sigma}$, as
\be
\tilde X^{\mu\nu\rho\sigma}=\sqrt{-\tilde g}e^{3\rho}\hat X^{\mu\nu\rho\sigma}-\frac{2\Theta}{\pi}
\epsilon^{\mu\nu\rho\sigma}.
\ee
As before, we calculate its inverse, defined as
\be
\tilde X^{\mu\nu\lambda\tau}(\tilde X^{-1})_{\lambda\tau\rho\sigma}=\delta^{\mu\nu}_{\rho\sigma}\;,
\ee
leading to the dual action (for comparison with subsection 3.2, note $a_\mu\rightarrow -B_\mu/2$, $\tilde X\rightarrow 
2C$, $\tilde X ^{-1}\rightarrow \tilde C/2$)
\be
S_{\rm dual}^\Theta=\int d^4x \left[-\frac{1}{4}f^{\mu\nu}(\tilde X_1^{-1})_{\mu\nu\rho\sigma}f^{\rho\sigma}\right]\;,
\ee
where ${(\tilde X_1^{-1})_{\mu\nu}}^{\rho\sigma}\equiv {\epsilon_{\mu\nu}}^{\lambda \tau}{(\tilde X^{-1})_{\lambda
\tau}}^{\eta\theta}{\epsilon_{\eta\theta}}^{\rho\sigma}$.

For the calculation of the inverse, we work to leading order in the Riemann tensor (or equivalently, in $E(g)_{\mu\nu\rho
\sigma}$ or, more practically, in $\lambda_{GB}$). We parametrize
\be
(\tilde X^{-1})_{\mu\nu\rho\sigma}=\frac{a}{\sqrt{-g}}g_{\mu[\rho}g_{\sigma]\nu}+\frac{b}{(-g)} 
\epsilon_{\mu\nu\rho\sigma}+\frac{c}{\sqrt{-g}}\lambda_{GB}L^2 E(g)_{\mu\nu\rho\sigma}. 
\ee

Imposing the inverse condition, we fix 
\be
a=\frac{e^{3\rho} Z}{(e^{3\rho}Z)^2+\left(\frac{4\Theta}{\pi}\right)^2}\;,\;\;\;
b=-\frac{\frac{2\Theta}{\pi}}{(e^{3\rho}Z)^2+\left(\frac{4\Theta}{\pi}\right)^2}\;,\;\;\;
c=-\frac{e^{3\rho}Y}{(e^{3\rho}Z)^2+\left(\frac{4\Theta}{\pi}\right)^2}.
\ee

Indeed, we have already calculated the inverse of the matrix without the Gauss-Bonnet term
in section 3.2, and it agrees with the above, 
considering that $K=e^{3\rho}Z$, whereas the extra GB term, assumed to be small, 
only comes with the opposite sign. 

Next, to write the currents, we only need to use the general formalism of the previous subsection, 
and to define $E(g)^{rxrx}\equiv r$, to obtain 
\bea
j^1&=&e^{3\rho}(Z+\lambda_{GB}L^2Yr)F^{01}+\frac{4\Theta}{\pi}F^{02}\nonumber\\
j^2&=&e^{3\rho}(Z+\lambda_{GB}L^2Yr)F^{02}-\frac{4\Theta}{\pi}F^{01}.
\eea

After the duality, we obtain instead
\bea
j^1&=&(a+c\lambda_{GB}L^2 r)f^{01}+2bf^{02}\cr
&=&\frac{e^{3\rho}(Z-\lambda_{GB}L^2Yr)}{e^{6\rho}(Z^{2}+(\frac{4\Theta}{\pi})^2)}f^{01}
+\frac{\frac{4\Theta}{\pi}}{e^{6\rho}Z^{2}+(\frac{4\Theta}{\pi})^2}f^{02}\cr
j^2&=&(a+c\lambda_{GB}L^2 r)f^{02}-2bf^{01}\cr
&=&\frac{e^{3\rho}(Z-\lambda_{GB}L^2Y r)}{e^{6\rho}(Z^{2}+(\frac{4\Theta}{\pi})^2)}f^{02}
-\frac{\frac{4\Theta}{\pi}}{e^{6\rho}Z^{2}+(\frac{4\Theta}{\pi})^2}f^{01}.
\eea

That means that the conductivities before the duality are 
\be
\sigma^{xx}=e^{3\rho}(Z+\lambda_{GB}L^2Yr)\;,\;\;\;
\sigma^{xy}=\frac{4\Theta}{\pi}\;,\label{sigmaxxxy}
\ee
and after the duality, they are 
\bea
\sigma_d^{xx}&=&\frac{e^{3\rho}(Z-\lambda_{GB}L^2Yr)}{e^{6\rho}Z^2+(\frac{4\Theta}{\pi})^2}\cr
\sigma_d^{xy}&=&-\frac{\frac{4\Theta}{\pi}}{e^{6\rho}Z^2+(\frac{4\Theta}{\pi})^2}.
\eea
This matches the formula for the inverse of (\ref{sigmaxxxy}), inverted as $\sigma^{-1}=-1/\sigma$, where 
$\sigma=\sigma^{xy}+i\sigma^{xx}$, but only up to corrections in $\Theta^2/e^{6\rho}Z^2$. More precisely, the
condition is
\be
\frac{\Theta^2}{e^{6\rho}Z^2}\ll \lambda_{GB}\frac{Y}{Z}r\ll 1.
\ee

\subsection{Effect of Weyl-Maxwell coupling and S-duality}

In this section we consider the effect on S-duality of 
a different type of quantum correction, explained first in \cite{Myers:2010pk}, in the presence of the theta term.

The general quantum-corrected Einstein-Maxwell action contains all possible covariant terms up to second order 
derivatives that preserve parity symmetry. Up to fourth order  in derivatives, it is possible to construct 15 covariant terms 
that preserve parity. However, by integration by parts and Bianchi identities, for both gauge and gravity identities 
($\nabla_{[a}F_{bc]}=0=R_{[abc]d}$), we are left with 8 independent terms,
\bea
S_{\rm qu.gr.,gen.}
&=&\int d^4x\sqrt{-g}\left[\alpha_1R^2+\alpha_2R_{\mu\nu}R^{\mu\nu}+\alpha_3(F^2)^2+\alpha_4F^4
+\alpha_5\nabla^\mu F_{\mu\nu}\nabla^\rho{F_\rho}^\nu\right.\cr
&&\left.+\alpha_6R_{\mu\nu\rho\sigma}F^{\mu\nu}F^{\rho\sigma}+\alpha_7R^{\mu\nu}F_{\mu\rho}{F_\nu}^\rho
+\alpha_8RF^2\right]
\eea
where $F^2=F_{\mu\nu}F^{\mu\nu}$ and $F^4={F^\mu}_\nu{F^\nu}_\rho{F^\rho}_\sigma{F^\sigma}_\mu$, 
$\alpha_i$ being unspecified couplings. 

The conductivity is not affected by the behavior of the terms of fourth order or higher in derivatives on the gauge field, 
but only up to second order in derivatives, so we only consider the ${\cal O}(F^2)$ which affect the transport properties of the 
dual field theory. 

From those terms with $\alpha_{6,7,8}$ which are second order in derivatives of the gauge fields, 
we can construct the generalized Maxwell term
\be
S_{\rm vector,qu.gr.}=\frac{1}{g^2_4}\int d^4x\sqrt{-g}
\left[-\frac{1}{4}F _{\mu\nu}F^{\mu\nu}+\gamma L^2C_{\mu\nu\rho\sigma}F^{\mu\nu}F^{\rho\sigma}\right]\;,
\ee
where $C_{\mu\nu\rho\sigma}$ is the Weyl tensor, which vanishes in pure AdS background. 
That means that at $T=0$ (zero temperature means AdS space), transport is unaffected by these corrections. Moreover, 
the planar AdS black hole is still a solution of the gravitational equations of motion, for the same reason. 
The equation of motion of the gauge field is now 
\be
\nabla_\mu[F^{\mu\nu}-4\gamma L^2C^{\mu\nu\rho\sigma}F_{\rho\sigma}]=0.
\ee

We can put the vector action into the general form (\ref{SM}), by considering 
\be
{X_{ab}}^{cd}=\delta_{ab}^{cd}-4\gamma L^2{C_{ab}}^{cd}.
\ee
Later on we will also want to add the theta term. 

As in the general quantum gravity example in the previous subsection, dualizing the action with $X^{\mu\nu\rho\sigma}$ 
amounts to just inverting $X^{\mu\nu\rho\sigma}$, which can be done to leading order in the curvature, here meaning 
to leading order in $\gamma$. We obtain
\be
{(X^{-1})_{\mu\nu}}^{\rho\sigma}=\delta_{\mu\nu}^{\rho\sigma}+4\gamma L^2{C_{\mu\nu}}^{\rho\sigma}
+{\cal O}(\gamma^2).
\ee

Then the master action 
\be
S_{\rm master}=
\int d^4x \left[-\frac{1}{4g^2}F_{\mu\nu}X^{\mu\nu\rho\sigma}F_{\rho\sigma}-\frac{1}{2}\epsilon^{\mu\nu\rho\sigma}
F_{\mu\nu}G_{\rho\sigma}\right]
\ee
gives the dual action 
\be
S_{\rm dual}=\int d^4x \left[-\frac{1}{4\tilde g^2}\hat X^{\mu\nu\rho\sigma}G_{\mu\nu}G_{\rho\sigma}\right]\/,
\ee
where $\tilde g^2=1/g^2$ (duality relation) and 
\be
{\hat X_{\mu\nu}}^{\rho\sigma}=-{\epsilon_{\mu\nu}}^{\lambda\tau}{(X^{-1})_{\lambda\tau}}^{\eta\theta}
{\epsilon_{\eta\theta}}^{\rho\sigma}.
\ee

The duality relation becomes
\be
F_{\mu\nu}=g^2{(\hat X^{-1})_{\mu\nu}}^{\lambda\tau}{\epsilon_{\lambda\tau}}^{\rho\sigma}G_{\rho\sigma}.
\ee

Finally, that means that the normal conductivity $\sigma=\sigma_{xx}$ 
is inverted, since in the original theory we have approximately  \cite{Myers:2010pk}
\be
\sigma=\frac{1}{g^2}(1+4\gamma)=\sigma_0(1+4\gamma)\;,
\ee
whereas in the dual theory we have 
\be
\sigma_d=\frac{1}{\tilde g^2}(1-4\gamma)=\frac{1}{\sigma_0}(1-4\gamma)\simeq \frac{1}{\sigma}.
\ee

We want now to introduce the theta term as well into the theory. 

First, we note that we can invert the original $X$ exactly, given the form of the planar AdS black hole background. 
Define $A,B\in \{tx,ty,tu,xy,xu,yu\}$, then $X$ becomes a diagonal six-by-six matrix
\begin{align}
{X_A}^B=\text{diag}(1+\alpha,1+\alpha,1-2\alpha,1-2\alpha,1+\alpha,1+\alpha)\;,
\end{align}
where $\alpha=4\gamma u^3$, ($u=\frac{r_o}{r}$).

Since $X$ is a diagonal matrix, then $X^{-1}$ is also a diagonal matrix, whose elements are the 
inverse of each element of $X$ diagonal matrix. Notice that $\alpha$ takes its maximum value at $u=1$, 
$\alpha_{max}=4\gamma$, which implies that $-\frac{1}{4}<\gamma<\frac{1}{8}$, 
in order for the inverse to exist in all the region outside the horizon.

Using the same notation and the background metric, ${\epsilon_{\mu\nu}}^{\rho\sigma}$ 
becomes the anti-diagonal $6\times 6$ matrix 
\begin{align}
{\epsilon_A}^B=
\begin{bmatrix}
 &  &  &  &  & \frac{r_0f}{L^2}\\ 
 &  &  &  & -\frac{r_0f}{L^2} & \\ 
 &  &  & \frac{L^2}{r_0} &  & \\ 
 &  & -\frac{r_0}{L^2} &  &  & \\ 
 & \frac{L^2}{r_0f} &  &  &  & \\ 
-\frac{L^2}{r_0f} &  &  &  &  & 
\end{bmatrix}.\label{epsilon}
\end{align}

Then the duality transformation becomes
\be
F_A=g^2_4{(X^{-1})_A}^B{\epsilon_B}^CG_C\;,\label{dualit}
\ee
which explicitly gives
\bea
F_{tx}&=&\frac{g^2_4}{1+\alpha}\frac{r_0f}{L^2}G_{yu}, \phantom{.....}
F_{ty}=-\frac{g^2_4}{1+\alpha}\frac{r_0f}{L^2}G_{xu},\nonumber\\
F_{tu}&=&\frac{g^2_4}{1-2\alpha}\frac{L^2}{r_0}G_{xy}, \phantom{.....}
F_{xy}=-\frac{g^2_4}{1-2\alpha}\frac{r_0}{L^2}G_{tu},\nonumber\\
F_{xu}&=&\frac{g^2_4}{1+\alpha}\frac{L^2}{r_0f}G_{ty}, \phantom{.....}
F_{yu}=-\frac{g^2_4}{1+\alpha}\frac{L^2}{r_0f}G_{tx}.\label{WeylFG}
\eea

Introducing the theta term, it means that we need to invert now 
\be
{\tilde X_{\mu\nu}}^{\rho\sigma}=\delta_{\mu\nu}^{\rho\sigma}-4\gamma {C_{\mu\nu}}^{\rho\sigma}
-\frac{2\Theta}{\pi}{\epsilon_{\mu\nu}}^{\rho\sigma}.
\ee
Note then that, for $|B A^{-1}|\ll 1$, we have the matrix relation
\be
(A+B)^{-1}\simeq A^{-1}-A^{-1}BA^{-1}.
\ee
Here $A$ refers to $X$ (the matrix at $\Theta=0$), and $B$ to the $\Theta$ term, so $A$ is diagonal, whereas $B$ is 
anti-diagonal. Also we denote, as in previous 
subsections, $1/g^2\equiv K$. Then the duality relation (\ref{dualit}) becomes
\be
F_A\simeq \frac{1}{K}\left[{X^{-1}_A}^B+\frac{2\Theta}{\pi}{X^{-1}_A}^C {\epsilon_C}^D {X^{-1}_D}^B\right]
{\epsilon_B}^E G_E.
\ee

This then gives explicitly
\bea
F_{tx}&\simeq &\frac{g^2}{1+\alpha}\frac{r_0f}{L^2}G_{yu}-\frac{2\Theta g^2}{\pi}\frac{1}
{(1+\alpha)^2}G_{tx}, \phantom{.....}F_{ty}\simeq-\frac{g^2}{1+\alpha}\frac{r_0f}{L^2}G_{xu}
-\frac{2\Theta g^2}{\pi}\frac{1}{(1+\alpha)^2}G_{ty},\cr
F_{tu}&\simeq &\frac{g^2}{1-2\alpha}\frac{L^2}{r_0}G_{xy}-\frac{2\Theta g^2}{\pi}
\frac{1}{(1-2\alpha)^2}G_{tu}, \phantom{...}F_{xy}\simeq -\frac{g^2}{1-2\alpha}\frac{r_0}{L^2}G_{tu}
-\frac{2\Theta g^2}{\pi}\frac{1}{(1-2\alpha)^2},\cr
F_{xu}&\simeq &\frac{g^2}{1+\alpha}\frac{L^2}{r_0f}G_{ty}-\frac{2\Theta g^2}{\pi}\frac{1}
{(1+\alpha)^2}G_{xu}, \phantom{.....}F_{yu}\simeq -\frac{g^2}{1+\alpha}\frac{L^2}{r_0f}G_{tx}
-\frac{2\Theta g^2}{\pi}\frac{1}{(1+\alpha)^2}G_{yu}.\cr
&&\label{WeylFGtheta}
\eea

However, for us the only relevant issue is the inversion of $\hat X$ which, for $\Theta/\pi\ll 1$ ($|AB^{-1}|\ll 1$)
and $\gamma \ll 1$ 
becomes approximately
\be
{(\hat X^{-1})_{\mu\nu}}^{\rho\sigma}\simeq {(X^{-1})_{\mu\nu}}^{\rho\sigma}
+\frac{2\Theta}{\pi}{\epsilon_{\mu\nu}}
^{\rho\sigma}.\label{inverseX}
\ee

For the currents, using the general formalism already developed, we find 
\bea
J^a&=& n_rX^{xrxr}F_{ra}+n_r\frac{4\Theta}{\pi}\epsilon^{ab}F_{0b}\;,\label{JJXY}
\eea
and using the general formula (\ref{frf0}) relating $F_{ra}$ with $F_{0a}$, we obtain 
\be
J^a=n_r \sqrt{-X^{xrxr}X^{txtx}}F_{0a}+n_r \frac{4\Theta}{\pi}\epsilon^{ab}F_{0b}\;,
\ee
where now 
\bea
X^{rxrx}(r_0)&=&g^{rr}g^{xx}{X_{ux}}^{ux}=(1+4\gamma)g^{rr}g^{xx}\cr
X^{txtx}(r_0)&=&g^{tt}g^{xx}{X_{tx}}^{tx}=(1+4\gamma)g^{tt}g^{xx}\;,
\eea
so the conductivities (before the duality relation) are 
\be
\sigma_{xx}=\frac{1+4\gamma}{g^2}\;,\;\;\;
\sigma_{xy}=\frac{4\Theta}{\pi}.
\ee

Because of the duality relation (\ref{inverseX}), after duality, the conductivities are 
\be
\sigma'_{xx}\simeq\frac{1-4\gamma}{\tilde g^2}\;,\;\;\;
\sigma'_{xy}\simeq -\frac{4\Theta}{\pi}\;,
\ee
which coincide indeed with the duality-transformed conductivities if $\gamma\ll 1, \Theta/\pi\ll 1$ and $\tilde g=1/g$.

\section{Fluid/gravity correspondence, membrane paradigm and S-duality effect on $\eta,\zeta$}

We now turn to a different kind of transport coefficient, namely the shear viscosity. In principle we can calculate the 
shear viscosity $\eta$ from the two-point function of the gravity perturbation, calculated holographically, by the use of one of the 
Kubo formulas,
\be
\eta(\omega,\vec{k})=\frac{iG^R_{T_{xy}T_{xy}}(\omega,\vec{k})}{\omega}.
\ee
This Kubo formula is derived by considering the variation of the viscous energy-momentum tensor (energy-momentum 
tensor for the fluid expanded up to one derivative acting on the relativistic fluid velocity $u_i$, $i=0,1,..,d-1$, with $d=3$
in the physical case),
\bea
T_{ij}&=&\rho u_i u_j +P(g_{ij}+u_i u_j)\cr
&&+2\eta\left[\frac{\nabla_i u_j +\nabla_j u_i}{2}-\frac{1}{d-1}(\nabla_k u^k)(g_{ij}+u_iu_j)\right]\cr
&&+\zeta (\nabla_k u^k)(g_{ij}+u_iu_j)\;,
\eea
with respect to a fluctuation in the $d$-dimensional metric $g_{ij}$, and equating with the response function, that involves
the retarded Green's function $G^R$.

Another way to holographically calculate $\eta$ is to compute the holographic energy-momentum tensor of the 
field theory from the gravitational action, with the correct holographic renormalization gravitational boundary terms, 
varied with respect to the metric at a surface $r=r_\infty$ near the boundary, 
\be
\langle T_{ij}\rangle _{FT}=\frac{\delta S_{\rm grav.,total}}{\delta g^{ij}_{\rm boundary}}\;,
\ee
where $g^{ij}_{\rm boundary}$ is the bulk metric $g^{\mu\nu}$ in the boundary directions and induced on the near-boundary 
surface $r=r_\infty$. This gives the result
\be
T^{\mu\nu}=\lim_{r_\infty\rightarrow \infty}\frac{[M_{\rm Pl}^{(d+1)}]^{d-1}r_\infty^{d-2}}{2}\left[K^{\mu\nu}-K 
g^{\mu\nu}-(d-1)g^{\mu\nu}-\frac{1}{d-2}\left(R^{\mu\nu}-\frac{1}{2}g^{\mu\nu}R\right)\right]\/,
\ee
which must be restricted to be in the boundary directions, and where
where $K^{\mu\nu}=g_{\mu\rho} \nabla^\rho n_\nu$ is the extrinsic curvature of the boundary surface. 
This implies
\be
\eta=\frac{[M_{\rm Pl}^{(d+1)}]^{d-1}}{2}\left(\frac{r_0}{L}\right)^{d-1}
=\frac{[M_{\rm Pl}^{(d+1)}]^{d-1}}{2}\left(\frac{4\pi}{d}TL\right)^{d-1}\;,\label{holosheard}
\ee
where $r_0$ (also called $r_H$ or $r_+$) is the position of the horizon,
which leads to the famous result $\eta/s=1/(4\pi)$. Usually one puts the radius of AdS to one, $L=1$.
Note that $M^8_{\rm Pl,10}L^8=\frac{N^2}{4\pi^5}$, and also equals $M^3_{\rm Pl,5}L^3/Vol_5
=M^3_{Pl,5}L^5/\pi^3$ in the $AdS_5\times S^5$ case,  where $Vol_5$ is the volume of the unit 5-sphere. 
In our physical case $d=3$, we obtain
\be
\eta=\frac{M^2_{\rm Pl}}{2}\left(\frac{r_0}{L}\right)^2=\frac{M^2_{\rm Pl}}{2}
\left(\frac{4\pi}{3}TL\right)^2.\label{holoshear}
\ee
In the literature, sometimes one puts both $L$ and $r_0$ to one.

However, in this section, we will follow \cite{Fischler:2015kro}, which extends the previous calculation of 
\cite{Price:1986yy}, that uses the original membrane paradigm formalism, in order 
to compute $\eta$ at the horizon, similarly to what we did in previous sections for the conductivities.
However, note that the original membrane paradigm of \cite{Price:1986yy} 
is slightly different (and gives somewhat different results)
for the shear viscosity than the paradigm used in AdS/CFT, for instance in \cite{Eling:2009sj}, and which 
gives the same result as the famous  KSS calculation \cite{Kovtun:2003wp}.

Instead of calculating at the boundary $r=r_\infty\rightarrow \infty$, we calculate the on-shell gravitational action 
with boundary terms at the horizon, and extract from the resulting boundary terms the energy-momentum tensor of the 
membrane (horizon), which we then put into the form of a fluid, and identify the transport coefficients. 

One considers a membrane ${\cal M}$, with normal unit vector $n^\mu$ ($n_\mu n^\mu=1$), and induced
metric on the 3-surface written as a 4-metric by
\be
h^{\mu\nu}=g^{\mu\nu}-n^\mu n^\nu\;,
\ee
and extrinsic curvature
\be
K_{\mu\nu}=h_\mu^\a h_\nu^\b \nabla_\a n_\b.
\ee
Finally the variation of the total on-shell gravitational action (including boundary terms) is
\be
\delta S_{\rm grav., total}^{\rm on-shell}=\frac{M^2_{\rm Pl}}{2}\int_{\cal M}d^3x \sqrt{-h} [h_{\mu\nu}K
-K_{\mu\nu}]\delta h^{\mu\nu}\;,
\ee
leading to a membrane energy-momentum tensor of 
\be
T_{\mu\nu}=M^2_{\rm Pl}\left.[K h_{\mu\nu}-K_{\mu\nu}]\right|_{\cal M}.\label{tmunum}
\ee

One considers a null generator of the horizon $l=\d/\d \bar t$, and coordinates on ${\cal M}$ as $x^i=(\bar t, x^A)$, 
where $A=1,2$ would correspond to $a=1,2$ on the boundary. We define the surface gravity at the horizon $\kappa$ by 
\be
l^\mu \nabla_\mu l^\nu =\kappa l^\nu.
\ee

In order not to confuse with $\nabla_\mu$, we call the covariant derivative using the 3-metric $h_{ij}$ by $D_i$,
understood as $D_i^{(3)}$. We similarly call the covariant derivative in $x^A$ (with metric $\gamma_{AB}$)
by $D_A$, understood as $D_A^{(2)}$. 

We can define coordinates $r$ such that the horizon is at $r=0$, define the coordinates to be comoving with $l$ so that $h_{0A}
=0$, and put the metric near the horizon in the form
\be
ds^2\simeq -r^2d\bar{t}^2+\frac{2r}{g_H}d\bar{t}dr+\gamma_{AB}\left(dx^A-\frac{\Omega^A(\bar{t}, x)}
{g_H}r^2d\bar{t}\right)\left(dx^B-\frac{\Omega^B(\bar{t}, x)}{g_H}r^2d\bar{t}\right)+{\cal O}(r^4).
\ee
so that $\gamma_{AB}=\gamma_{AB}(\bar{t}, r, x)$ is the metric on the horizon surface  $r=0$. 

We can also define, in these comoving coordinates, 
\be
\frac{\partial \gamma_{AB}}{\partial \bar{t}}= 2\sigma^{H}_{AB}+\theta_H\gamma_{AB}\;,
\ee
where the horizon shear $\sigma^H_{AB}$ and horizon expansion $\theta_H$ are 
\bea
\sigma^H_{AB}&=& \theta_{AB}-\frac{1}{2}\gamma_{AB}\theta_H\cr
\theta_H&=&\gamma^{AB}\theta_{AB}=\frac{\d}{\d\bar t}\ln \sqrt{\gamma}\cr
\theta_{AB}&=&D_A^{(2)} l_B.
\eea

Then we replace the horizon with the stretched horizon (membrane) at $r=\epsilon$, and define a fluid living almost at
rest in the comoving coordinates, with 
\be
u^{\bar t}=\frac{1}{\epsilon}\;,\;\;\;
u^A={\cal O}(\epsilon)\;,
\ee
which is $u^i=U^i+{\cal O}(\epsilon)$, where 
\be
U=-rd\bar t+\frac{dr}{\kappa}\Rightarrow U^{\bar t}=\frac{1}{r}\;,\; U^r=0\;,\; U^A=-\frac{\Omega^Ar}{\kappa}.
\ee
One obtains for $\epsilon\rightarrow 0$ that
\be
D_i u^i=\frac{1}{\epsilon}\theta_H.
\ee

Then one calculates the extrinsic curvature components
\bea
&& K_{00}(r=\epsilon)=-\kappa \epsilon +{\cal O}(\epsilon)\;,\;\;\;K_{0A}(r=\epsilon)={\cal O}(\epsilon)\;,\;\;\;
K_{AB}(r=\epsilon)=\frac{1}{\epsilon}\theta_{AB}\Rightarrow\cr
&&K=K_{ij}h^{ij}=\frac{1}{\epsilon}(\kappa+\theta_H)+{\cal O}(\epsilon)\;,
\eea
which finally allow us to define the energy-momentum tensor of the stretched horizon membrane. 

This is then identified with a fluid energy-momentum tensor, the sum of an ideal term and a viscous term $\pi^{ij}$,
\bea
T^{ij}&=&(\rho+P)u^i u^j +Ph^{ij}+\pi^{ij}\cr
\pi^{ij}&=&P^{ik}P^{jl}(\eta f_{kl}+\zeta h_{kl}D_m u^m)\cr
P^{ij}&=& h^{ij}+u^i u^j\cr
f_{ij}&=& D_i u_j +D_j u_i -h_{ij}D_k u^k\;,
\eea
where $P^{ij}$ is a projector.

We now calculate
\be
f_{00}=f_{0A}={\cal O}(\epsilon)\;,\;\;\;
f_{AB}=\frac{1}{\epsilon}\sigma^H_{AB}.
\ee

We first note that, since $u_i P^{ij}=0$, we obtain the projections
\be
\rho=u_i u_i T^{ij}\;,\;\;\;
P_{ij}T^{ij}=2(P-\zeta D_i u^i)\;,
\ee 
which allows us to separate the non-shear viscosity terms (and leave the remaining ones as shear viscosity terms). 
By equating the energy-momentum tensor with the one in (\ref{tmunum}), we obtain 
\be
\rho=-M^2_{\rm Pl}(K+K_{ij}u^i u^j)\;,\;\;\;
2(p-\zeta D_i U^i)=M^2_{\rm Pl}(K-K_{ij}u^i u^j).
\ee

Finally, this allows us to split the membrane energy-momentum tensor as
\bea
T^{ij}&=&M_{\rm Pl}^2\left.\left[(-K-K_{kl}u^ku^l)u^iu^j+\frac{1}{2}(K-K_{kl}u^ku^l)P^{ij}\right]
\right|_{\mathcal{M}}\nonumber\\
&&+M_{\rm Pl}^2\left.\left[\frac{1}{2}KP^{ij}-K^{ij}+(K_{kl}u^ku^l)\left(u^iu^j+\frac{1}{2}P^{ij}\right)\right]\right|
_{\mathcal{M}}.\label{Tfluid1}
\eea
Then, by  identifying the two terms on the first line as $\rho$ term and $P-\zeta D_i u^i$ term, and the remaining ones, 
on the second line, as $\eta$ terms, allows us to calculate
\be
\rho=-\frac{M_{\rm Pl}^2}{\epsilon}\theta_H, \phantom{....} P=\frac{M_{\rm Pl}^2}{\epsilon}\kappa, 
\phantom{.....} \zeta=-\frac{M_{\rm Pl}^2}{2}, \phantom{....} \eta=\frac{M_{\rm Pl}^2}{2}.
\ee

This result is however problematic, because the bulk viscosity $\zeta$ is negative, and one also obtains an unphysical 
entropy current, and moreover the shear viscosity 
is different than the holographic result (\ref{holoshear}). 
But in  \cite{Eling:2009sj} an update of this standard membrane paradigm calculation, tailor made for the 
kind of AdS black hole solution relevant to gravity duals was made, which does obtain (\ref{holoshear}).

The idea of writing the energy-momentum tensor of a general quantum field theory in the form of one of a fluid  was also 
used in \cite{Nastase:2015ljb,Endlich:2010hf,Dubovsky:2011sj,Berezhiani:2016dne}, though there it was mostly 
applied to a scalar field. The procedure is however general: any quantum field theory, in particular if it is strongly coupled, 
can be written in the hydrodynamics expansion, by identifying the energy-momentum tensor of the system with the 
one of a fluid. But as noted also in these papers, there is a degree of ambiguity when one does that, both because 
there is an ambiguity in what one considers to be the velocity of the fluid $u^i$, and because usually there is some 
freedom in defining what terms in the energy-momentum tensor map to what terms in the fluid expansion. 

This kind of ambiguity was used in \cite{Eling:2009sj} where it was shown that one can fix the ambiguity in a 
physical way. One considers first a zeroth order metric, the metric of a planar AdS black hole boosted by a 
relativistic velocity $u^i$, which is (at AdS radius $L=1$)
\be
ds^2=-2u_i dx^i dr +\frac{r_0^d}{r^{d-2}}u_i u_j dx^i dx^j +\eta_{ij}dx^i dx^j.
\ee
The normal vector to the horizon is $l^\mu: (l^r=0,l^i=u^i)$. Then one introduces spacetime dependence of
$u^i$ and the temperature $T$, turning them into fields, described as $u^i(\epsilon x^j)$ and $T(\epsilon x^j)$. 
The metric will be modified by ${\cal O}(\epsilon)$ terms as well, and the previous expansion in $\epsilon$ (defining 
the stretched horizon) is replaced by the new one defined here, with the same meaning. 
Moreover, to fix the ambiguity of the stress-tensor, we choose the Landau frame for the viscous part of the 
energy-momentum tensor, as 
\be
u_i \pi^{ij}=0\;,
\ee
and is undestood to be imposed on the membrane (horizon) tensor. Finally, this gives the correct result (\ref{holosheard}), 
which corrects $\eta$ by $(r_0/L)^{d-1}$. We note then that the velocity field is different with respect to the 
previous case, which accounts for the different transport parameters when writing $T_{ij}$.

But we now want to study the effect of adding a vector field with a theta term, and a scalar field, and the effect 
of S-duality on the result. As we have stated, we will take the point of view 
of  \cite{Nastase:2015ljb,Endlich:2010hf,Dubovsky:2011sj,Berezhiani:2016dne} and add the energy-momentum tensor 
of the vector or scalar field to the previous one, and try to put it also in the form of a viscous fluid one. More precisely, since 
we are using AdS/CFT together with the membrane paradigm, we will proceed in the same way as in the case of the 
pure gravity part: we will consider only the boundary term, {\em at the horizon}, of the bulk vector+scalar action, dual 
to the dynamics of vector and scalar operators in the strongly coupled field theory. 

Considering fields $\phi_I$ and sources for them at the''boundary'' = stretched horizon, 
\be
S_{\rm fields}=\int_Ad^{d+1}x\sqrt{-g}\mathcal{L}(\phi_I,\nabla_{\mu }\phi_I)
+\sum_I \int_{\mathcal{M}}d^dx\sqrt{-h}\mathcal{J}^I_{\mathcal{M}}\phi_I\;,
\ee
where $h$ is the determinant of the induced metric on the stretched horizon $\mathcal{M}$, and the boundary term 
is needed to cancel the boundary term obtained in the bulk by partial integration.

For the case of a vector field, the relevant boundary term is written as
\be
S_{\rm surf}[\phi_I]=\int _{\mathcal{M}}d^dx\sqrt{-h}\mathcal{J}^{\mu}_{\mathcal{M}}A_{\mu}\;,
\ee
and from it we derive the current at the ``boundary'',
\be
\mathcal{J}^{\mu}_{\mathcal{M}}=\left.n_{\nu}\frac{\partial \mathcal{L}}{\partial(\nabla_{\nu}
A_{\mu})}\right|_{\mathcal{M}}.
\ee
For a  bulk term 
\be
S_{\rm vector, bulk}=\int d^4x \sqrt{-g}\left[-\frac{1}{4}F_{\mu\nu}F^{\mu\nu}
+\frac{\Theta}{2\pi}F_{\mu\nu}\tilde F^{\mu\nu}\right]\;,\
\ee
we obtain the horizon current
\be
\mathcal{J}^{\mu}_{\mathcal{M}}=Kn_{\nu}F^{\mu\nu}-\frac{2\Theta}{\pi}n_{\nu}\tilde  F^{\mu\nu}.
\ee

Now we rewrite the boundary term using this form of the current as 
\bea
S^{A_{\mu}}_{\rm surf}&=&\int_{\mathcal{M}}d^3x\sqrt{-h}\left[Kn_{\nu}F^{\mu\nu}-\frac{\Theta}{\pi}n_{\nu}\tilde
F^{\mu\nu}\right]A_{\mu}\cr
&=&\int_{\mathcal{M}}d^3x\sqrt{-h}\frac{Kf_{\mu\nu}}{2}\left[F^{\mu\nu}-\frac{\Theta}{\pi K}\tilde F^{\mu\nu}\right].
\eea
Here the field strength in the radial directions is $f_{\mu\nu}=n_\mu A_\nu-n_\nu A_\mu$. Varying this action with respect
to the boundary metric $h^{ij}$, we get
\be
T_{ij}^{A_\mu}\equiv  -\frac{2}{\sqrt{-h}}\frac{\delta S^{A_\mu}_{\rm surf}}{\delta h^{ij}}=
-\frac{K}{4}h_{ij}f_{\mu\nu}F^{\mu\nu}+Kf_{i\mu}{F_j}^{\mu}.\label{Tafield}
\ee
Note that the topological theta term doesn't contribute, since the variation with respect to the metric is zero for any topological 
contribution.

We rewrite \eqref{Tafield} similarly to what was done for  \eqref{Tfluid1},
\be
T^{A_{\mu}}_{ij}=K\left(-\frac{1}{4}h_{ij}f_{lk}F^{lk}+f_{il}{F_j}^l\right)+K\frac{n_r}{g_{rr}}\left(-
\frac{1}{2}h_{ij}A_l\partial_rA^l+A_i\partial_rA_j\right).
\ee
Since $n_r=\sqrt{g_{rr}}$, we first define
\bea
\mathcal{F}&=&f_{mn}F^{mn}, \phantom{.....} \mathcal{F}_{ij}=f_{il}{F_j}^l.\nonumber\\
\mathcal{A}&=&\frac{1}{\sqrt{g_{rr}}}A_l\partial_rA^l, \phantom{.....} \mathcal{A}_{ij}=\frac{1}{\sqrt{g_{rr}}}A_i\partial_rA_j\;,\label{AFcal} 
\eea
and then rewrite the energy-momentum tensor as 
\be
T^{A_{\mu}}_{ij}=K\left[\left(-\frac{h_{ij}}{4}\mathcal{F}+\mathcal{F}_{ij}\right)+\left(-\frac{h_{ij}}{2}
\mathcal{A}+\mathcal{A}_{ij}\right)\right].
\ee

In turn, this can again be split in parts corresponding to the viscous fluid energy-momentum tensor, 
as in the pure gravity case. We first note that
\bea
u^iu^jT^{A_{\mu}}_{ij}&=&K\left[\frac{\mathcal{F}}{4}+u^iu^j\mathcal{F}_{ij}+\frac{\mathcal{A}}{2}
+u^iu^j\mathcal{A}_{ij}\right]\cr
\frac{1}{2}P^{ij}T^{A_{\mu}}_{ij}&=& K\left[-\frac{\mathcal{F}}{4}
-\frac{\mathcal{A}}{2}+\frac{P^{ij}}{2}(\mathcal{F}_{ij}+\mathcal{A}_{ij})\right].
\eea
Finally, that allows us to write the energy-momentum tensor as 
\bea
T_{A_\mu}^{lk}&=&\left(\frac{\mathcal{F}}{4}+\frac{\mathcal{A}}{2}+u^iu^j(\mathcal{F}_{ij}
+\mathcal{A}_{ij})\right)u^lu^k\cr
&&+\left(-\frac{\mathcal{F}}{4}-\frac{\mathcal{A}}{2}+\frac{1}{2}P^{ij}(\mathcal{F}_{ij}+\mathcal{A}_{ij})\right)
P^{lk}-\eta P^{lm}P^{kn}f_{mn}\;,
\eea
where the last term is the one that contains the viscosity, and contains the remaining terms in the energy-momentum tensor,
\be
-\eta P^{lm}P^{kn}f_{mn}=\mathcal{G}^{lk}-\frac{1}{2}P^{lk}\mathcal{G}
-u^iu^j\mathcal{G}_{ij}\left(u^lu^k+\frac{1}{2}P^{lk}\right)\;,
\ee
where we have defined 
\be
\mathcal{G}_{ij}\equiv \mathcal{F}_{ij}+\mathcal{A}_{ij}.
\ee

We see that the formalism allows for adding the vector field contribution to the standard gravity contribution in the 
same way, by simply (and formally) adding an extra term to the extrinsic curvature of the stretched horizon surface, 
by defining
\be
\mathcal{K}_{ij}=K_{ij}+\mathcal{G}_{ij}.
\ee

We can in fact extend the same analysis to the case of a scalar field with a canonical kinetic term. In this case, the surface
term is 
\be
S^{\phi}_{\rm surf}=-\frac{1}{2}\int d^3x\sqrt{-h}[h^{\mu\nu}n_{\mu}\phi\partial_{\nu}\phi] \;,
\ee
leading to an energy-momentum term contribution of 
\be
T_{ij}^\phi=n_i\phi\partial_j\phi-\frac{1}{2}h_{ij}n^{\mu}\phi\partial_{\mu}\phi.
\ee

Defining 
\be
\mathcal{P}=n^l\phi\partial_l\phi,\phantom{.....} \mathcal{P}_{ij}=n_i\phi\partial_j\phi\/,
\ee
we can repeat the above procedure to write the energy-momentum tensor contribution $T_{ij}^\phi$ as 
a fluid one, isolating the shear viscosity one as 
\be
-\eta (PPf)^{lk}= \mathcal{P}^{lk}-\frac{1}{2}P^{lk}\mathcal{P}-u^iu^j\mathcal{P}_{ij}
\left(u^lu^k+\frac{1}{2}P^{lk}\right).
\ee

We see then that the contribution to the energy-momentum tensor for the scalar field can be reproduced again by 
simply adding an extra term to the extrinsic curvature of the stretched horizon surface, obtaining in total 
\be
\mathcal{K}_{ij}=K_{ij}+\mathcal{G}_{ij}+\mathcal{P}_{ij}.
\ee

The conclusion is then that the matter contribution doesn't modify the value of the shear viscosity. Of course, here we 
followed the old membrane paradigm formulation, leading to the wrong value of the shear viscosity, $\eta=M^2_{\rm Pl}/2$, 
but since the matter contribution is simply encapsulated by adding an extra term to the extrinsic curvature, when repeating 
the exact AdS/CFT procedure leading to the correct $\eta/s=1/(4\pi)$ in this case, nothing is changed, and the 
value of $\eta$ (and $\eta/s$) is unmodified. 

Moreover, we have seen that in fact the theta term didn't even modify the energy-momentum tensor itself, let alone $\eta$. 
Therefore we can say that S-duality, which acts on (the coefficients of) the matter action for the vector field, will have 
no effect on $\eta$.

\section{Conclusions}

In this paper we have considered the action of particle-vortex duality and the effect of theta terms, from the point of 
view of the AdS/CMT correspondence. 

We have defined the action of particle-vortex duality on 2+1 dimensional field theories for a scalar coupled to a 
Maxwell field, with scalar function $K$ and $\Theta$ term (Chern-Simons term). We have calculated the action of 
particle-vortex duality on $K$ and $\Theta$, and the corresponding action on conductivities of the field theory, amounting 
to $\sigma'=-1/\sigma$, with $\sigma=\sigma_{xy}+i\sigma_{xx}$. Then considering an AdS/CMT ansatz for a 
3+1 dimensional gravitational theory with a black hole solution with a horizon, we have calculated the action of 
S-duality on a Maxwell field with scalar function $K$ and theta term (topological term with $\Theta$), finding that 
it reduces to the same action on $K$ and $\Theta$, and moreover, it amounts to the same relation $\sigma'=-1/\sigma$ 
for the conductivity of the horizon, evaluated in a membrane type paradigm. Moreover, the relation between the 3+1 
dimensional and the 2+1 dimensional cases is consistent with AdS/CFT, as we have shown explicitly. 

Quantum gravity corrections in the gravitational bulk were also considered, finding that the presence of Gauss-Bonnet-Maxwell
corrections doesn't change the duality relation for the conductivity, as long as second order corrections in $\Theta$ are
negligible, and smaller than Gauss-Bonnet corrections. Moreover, we have  found that the Weyl-Maxwell coupling 
(standing in for quantum gravity corrections) also doesn't change the form of the duality relation for the conductivity. 
A membrane paradigm calculation of the shear viscosity, obtained by writing a boundary energy-momentum tensor 
at the horizon for gravity, and then putting it into fluid form, showed that adding vectors and scalars, thus modifying the 
boundary energy-momentum tensor, nevertheless has no effect on $\eta,\zeta$, and thus S-duality doesn't affect them.

\section*{Acknowledgements}

We thank Aristomenis Donos for useful discussions. The work of HN is supported in part by CNPq grant 304006/2016-5 and FAPESP grant 2014/18634-9. HN would also
like to thank the ICTP-SAIFR for their support through FAPESP grant 2016/01343-7.
The work of LA is supported by Capes grant 2017/19046-1.

\appendix

\section{Particle-vortex duality review}

In this appendix we review the particle-vortex duality derived first in \cite{Burgess:2000kj} and then clarified and 
extended in \cite{Murugan:2014sfa}.

\subsection{Burgess-Dolan form of particle-vortex duality}

We review here the work of Burgess and Dolan in  \cite{Burgess:2000kj}.

Defining $\phi$ as the phase angle of the complex scalar field $\Phi=|\Phi|e^{-i\phi}$, in the presence of vortices 
we have
\be
\phi(\theta+2\pi)=\phi(\theta)+2\pi \sum_aN_a\;,
\ee
where $N_a$ is the \emph{vorticity} or \emph{winding number} of vortex $a$. We then split 
$\phi=\omega+\varphi$, where $\varphi$ is the vortex-free part, satisfying periodic boundary conditions, 
$\varphi(\theta+2\pi)=\varphi(\theta)$, and 
$\omega(x)$ is an explicit muit-vortex solution, containing all the nontrivial part,
\be\label{A5}
\omega(x)=\sum_a N_a \arctan \left(\frac{x^1-y^1_a}{x^2-y^2_a}\right)\equiv \sum_a N_a \theta_a.
\ee
Here we have defined
\be
\frac{x^1-y_a^1}{x^2-y^2_a}=\tan \theta_a
\ee 
as the angle of rotation around a particular vortex. We calculate the gradient of the vortex part, 
\be
v_\mu\equiv \d_\mu \omega
=\sum_a N_a\frac{1}{1+\tan^2\theta_a}\d_\mu \tan \theta_a=\sum_a N_a \d_\mu \theta_a\;,
\ee
which gives the vortex current
\be
j^\mu(t) = j^\mu_{\rm vortex}(t)=\sum_a N_a\ \dot{y}_a^\mu\ \delta[x-y_a(t)]\label{vortexcurrent}
\ee
from 
\be
\epsilon^{\mu\nu\rho}b_\mu\d_\nu v_\rho = b_\mu \sum_a N_a\ \epsilon^{\mu\nu\rho}\d_\nu\d_\rho \theta_a = 2\pi b_\mu\sum_a N_a\ \dot{y}_a^\mu\ \delta[x - y_a(t)] = 2\pi b_\mu j^\mu(t)\;.
\ee
On the other hand, the electric (particle) current associated with a  canonical complex scalar field is 
\be
j_\mu =\frac{ie}{2}\left[\Phi^\dagger \d_\mu \Phi -(\d_\mu \Phi^\dagger)\Phi\right]=e|\Phi|^2\d_\mu\theta\;.\label{scalarcurrent}
\ee

For a complex scalar fiedl $\Phi$ coupled with a Chern-Simons gauge field $a$ and an external gauge field $A$, having 
an arbitrary Higgs potential depending only on $|\Phi|^2$, 
\be
S = -\frac{1}{2}\int\left[\left[(i\d_\mu -e\tilde a_\mu )\Phi\right]^\dagger \left[(i\d^\mu -e\tilde a^\mu)\Phi\right] + \frac{\pi e^2}{\theta}\epsilon^{\mu\nu\rho}a_\mu \d_\nu a_\rho\right] + S_{\rm int}\left[|\Phi|^2\right]\;,
\ee
where $\tilde a\equiv a+A$, we then split the field in an absolute value, a smooth phase $\theta$ and a vortex part, according 
to 
\bea
\Phi(\vec{r})&=&\Phi_0(\vec{r})e^{-i\theta(\vec{r})}v(\vec{r})\cr
v(\vec{r})&=&\exp\left[\frac{2\pi i}{q_\phi}\sum_a N_a \arctan \left(\frac{x^1-y^1_a}{x^2-y^2_a}\right)\right]\;.
\eea

Then the action becomes 
\bea
S^a[\Phi_0,\theta,a,A]
&=& -\frac{1}{2}\int\left[(\d_\mu \Phi_0)^2 + e^2\Phi_0^2\tilde a_\mu \tilde a^\mu 
+ \frac{1}{e^2\Phi_0^2}j_\mu j^\mu - 2\tilde a_\mu j^\mu \right]\cr 
&&- \frac{\pi e^2}{2\theta}\int\epsilon^{\mu\nu\rho}a_\mu \d_\nu a_\rho
 + S_{\rm int}[\Phi_0^2]\;,
\eea
and the particle current splits into a smooth and a vortex contribution, 
\be
j_\mu =e\Phi_0^2(\d_\mu \theta + iv^*\d_\mu v)\;.
\ee

We define $\lambda_\mu=\d_\mu \theta$, and then make $\lambda_\mu$ independent, but subject to the constraint
$\epsilon^{\mu\nu\rho}\d_\nu \lambda_\rho=0$ imposed with a Lagrange multiplier $\tilde b_\mu$, with relevant 
path integration (over a master action)
\be
\int {\cal D}\lambda_\mu {\cal D}\tilde b_\mu \exp\left[-\frac{i}{2}\int(\lambda_\mu + iv^*\d_\mu v - e\tilde a _\mu)^2\Phi_0^2 + \epsilon^{\mu\nu\rho}\tilde b_\mu \d_\nu \lambda_\rho\right]\;.
\ee
Doing instead the integration over $\lambda_\mu$ first, we obtain the dual action in terms of the Lagrange multiplier as a
dual field, 
\bea
S^b[\Phi_0,A,a,\tilde b] &=& \int\left[-\frac{1}{4e^2\Phi_0^2}\tilde f^{(b)}_{\mu\nu}\tilde f^{(b)\mu\nu}+\tilde j^\mu \tilde b_\mu-\epsilon^{\mu\nu\rho}\tilde a_\rho
\d_\nu \tilde b_\rho-\frac{\pi e^2}{2\theta}\epsilon^{\mu\nu\rho}a_\mu \d_\nu a_\rho\right]\cr
&& -\frac{1}{2}\int \d_\mu\Phi_0\d^\mu \Phi_0 + S_{\rm int}'\left[\Phi_0^2\right]\;,
\eea
where the dual field strength is $\tilde f^{(b)}_{\mu\nu}=\d_\mu \tilde b_\nu -\d_\nu \tilde b_\mu$, and 
\be
\tilde j^\mu =\frac{i}{e}\epsilon^{\mu\nu\rho}\d_\nu v^*\d_\rho v\;,
\ee
which before was part of the electric (particle) current, 
is now the {\em vortex} current.

\subsection{Exact duality in the path integral}

We next review the refinement of the duality in  \cite{Murugan:2014sfa} by writing the duality completely at the level 
of the path integral, in a generic theory. 

Consider a complex scalar $\Phi$ coupled to a $U(1) $ gauge field through the action 
\be
S=\int d^3x \left[-\frac{1}{2}|D_\mu \Phi|^2-V(|\Phi|)-\frac{1}{4}F_{\mu\nu}^2\right]\;,
\ee
where $F_{\mu\nu}=\d_\mu a_\nu -\d_\nu a_\mu$ and $D_\mu \Phi=\d_\mu \Phi-iea_\mu \Phi$. The path integral for 
this action is done over $a_\mu, \Phi_0$ and $\theta$, where the scalar is split as $\Phi=\Phi_0 e^{i\theta}$. For a vortex 
solution, $\Phi_0(r) e^{i\theta(\a)}$, 
with $(r,\a)$ the polar coordinates in 2 dimensions, and $\theta(\a)=N\a$. 

We then split the phase of $\Phi$ into a smooth part (with no vortices) and a vortex part, 
\be
\theta=\theta_{\rm smooth}+\theta_{\rm vortex}\;,
\ee
so that $\epsilon^{ab}\d_a \d_\b \theta_{\rm smooth}=0$, but $\epsilon^{ab}\d_a \d_b \theta_{\rm vortex}\neq 0$.
Under this split, the action becomes
\bea
S&=&-\frac{1}{2}\int d^3x \left[(\d_\mu\Phi_0)^2+(\d_\mu \theta_{\rm smooth}+\d_\mu \theta_{\rm vortex}+ea_\mu)^2
\Phi_0^2\right]\cr
&&-\int d^3x\left[V(\Phi_0)+\frac{1}{4}F_{\mu\nu}^2\right].
\eea

We next replace $\d_\mu\theta$ with an independent variable $\lambda_\mu$, imposing the flatness of its curvature 
by $\epsilon^{\mu\nu\rho}\d_\nu \lambda_\rho=0$, with Lagrange multipliers $b_\mu$, which leads to the master
action 
\bea
S_{\rm master}&=&\int d^3x \left[-\frac{1}{2}(\d_\mu\Phi_0)^2-\frac{1}{2}(\lambda_{\rm \mu,smooth}+\lambda_{\rm 
\mu, vortex}+ea_\mu)^2\Phi_0^2+\epsilon^{\mu\nu\rho}b_\mu \d_\nu \lambda_{\rm \rho, smooth}\right.\cr
&&\left. -V(\Phi_0)-\frac{1}{4}F_{\mu\nu}^2 \right].
\eea
The path integral for this master action is done over $\lambda_\mu, b_\mu, a_\mu, \Phi_0$.

We check that by varying with respect to $b_\mu$ or (since the action is linear in it) by path integrating over it, 
we obtain that $\lambda_\mu$ is the $\d_\mu$ of something, leading 
back to the original action. If we vary with respect to $\lambda_{\rm \mu, smooth}$ 
instead (or rather, do the path integration over 
$\lambda_\mu$, as this is a simple quadratic one), we obtain 
\be
(\lambda_\mu+ea_\mu)\Phi_0^2=e\epsilon^{\mu\nu\rho}\d_\nu b_\rho\;,
\ee
and by substituting in the master action (or rather, doing the path integration over $\lambda_\mu$), we obtain 
the dual action, 
\be
S_{\rm dual}=\int d^3x \left[-\frac{(f_{\mu\nu}^b)^2}{4\Phi_0^2}-\frac{1}{2}(\d_\mu \Phi_0)^2-e \epsilon^{\mu\nu
\rho}b_\mu \d_\nu a_\rho-\frac{2\pi}{e}b_\mu j^\mu _{\rm vortex}-V(\Phi_0)-\frac{1}{4}F_{\mu\nu}^2\right]\;.
\ee

Then the duality exchanges the electric current, 
\be
j_\mu=e\Phi_0^2\d_\mu\theta
\ee
with the vortex current
\be
j^\mu_{\rm vortex}=\frac{e}{2\pi}\epsilon^{\mu\nu\rho}\d_\nu \d_\rho \theta=\frac{1}{2\pi \Phi_0^2}\d_\nu j_\rho\;,
\ee
and exchanges the phase $\theta$ of the scalar field $\Phi$ with the gauge field $b_\mu$, by
\be
\d_\mu \theta+ea_\mu=\frac{1}{\Phi_0^2}\epsilon^{\mu\nu\rho}\d_\nu b_\rho\;,
\ee
which is nothing but Poincar\'{e} duality in 3 dimensions. 

This particle-vortex duality is also like an S-duality (strong/weak duality) in the sense that it inverts the coupling. 
Indeed, for the scalar $\theta$ in the original action, $\Phi_0^2$ acts as the coupling factor $1/g^2$, whereas for the 
dual action, $1/\Phi_0^2$ acts as the dual coupling factor $1/\tilde g^2$, leading to $\tilde g=1/g$.

We note that this duality has some remarkable similarities with the Mukhi-Papageorgakis Higgs mechanism
\cite{Mukhi:2008ux} in 2+1 
dimensional gauge theories with Chern-Simons terms, by which a Chern-Simons gauge field (with no dynamical degrees
of freedom) eats a real scalar (the phase of a complex scalar field) and becomes Maxwell, which is a dynamical field with 
one degree of freedom. 

In it, a complex scalar $\Psi$ coupled to a Chern-Simons gauge field $a_\mu$ with action 
\be
 S = -\int d^3x\ \left[\frac{k}{2\pi} \epsilon^{\mu\nu\rho}a_\mu \d_\nu \tilde a_\rho 
 + \frac{1}{2}|(\d_\mu -iea_\mu)\Psi|^2 
  + V(|\Psi|^2)
\right]\,,\label{cs}
\ee
and with minimum of the potential (vacuum solution) at $\Psi=b$, is expanded around the vacuum as
\be
  \Psi = (b + \delta \psi)e^{-i\delta\theta};\;\;\;\; \delta \theta = \theta_{\rm smooth} + \theta_{\rm vortex}\, ,
\ee
leading to a perturbative action
\be
  S = -\int d^3x\ \left[\frac{k}{2\pi} \epsilon^{\mu\nu\rho}a_\mu \d_\nu \tilde a_\rho+\frac{1}{2}(\d_\mu \delta \psi)^2 +   
  \frac{1}{2}(\d_\mu \theta_{\rm smooth} + \d_\mu \theta_{\rm vortex} + ea_\mu)^2b^2 + \ldots\right]\,.
\ee
Redefining the gauge field so that it ``eats'' the real Higgs field $\theta$, by
\begin{equation}\label{eq.MPmech}
ea_\mu + \d_\mu\theta_{\rm smooth} + \d_\mu \theta_{\rm vortex} = ea'_\mu\,,
\end{equation} 
and solving for  (integrating out in the path integral) $\theta$ and $a'_\mu$, we get the perturbative action
\be
S = \int d^3x\ \left[-\frac{ k^2}{16\pi^2b^2}(\tilde f_{\mu\nu})^2 - \frac{1}{2}(\d_\mu \delta\psi)^2 + \frac{k}{e} j^{\mu}_{\rm vortex}\tilde a_\mu + \ldots \right]\,,\label{higgs}
\ee
where $\tilde f_{\mu\nu}=\d_\mu \tilde a_\nu -\d_\nu \tilde a_\mu$. The solution for $a'_\mu$ provides a 
relation similar to particle-vortex duality, 
\be
  a^\mu+\frac{1}{e}\d^\mu\delta \theta = a'^\mu = -\frac{k}{2\pi b^2}\epsilon^{\mu\nu\rho}\d_\nu \tilde a_\rho\,.
\ee

It is then possible to combine particle-vortex duality with the Mukhi-Papageorgakis Higgs mechanism to obtain a symmetric
kind of particle-vortex duality that relates two scalars coupled with two gauge fields with a similar dual action, thus obtaining 
a kind of self-duality.

\bibliography{PVcond}

\providecommand{\href}[2]{#2}\begingroup\raggedright\begin{thebibliography}{10}

\bibitem{Zee:2003mt}
A.~Zee, {\em {Quantum field theory in a nutshell}}.
\newblock
2003.
\newblock

\bibitem{Murugan:2016zal}
J.~Murugan and H.~Nastase, ``{Particle-vortex duality in topological insulators
  and superconductors},'' \href{http://dx.doi.org/10.1007/JHEP05(2017)159}{{\em
  JHEP} {\bf 05} (2017)  159},
\href{http://arxiv.org/abs/1606.01912}{{\tt arXiv:1606.01912 [hep-th]}}.

\bibitem{Karch:2016sxi}
A.~Karch and D.~Tong, ``{Particle-Vortex Duality from 3d Bosonization},''
  \href{http://dx.doi.org/10.1103/PhysRevX.6.031043}{{\em Phys. Rev.} {\bf X6}
  (2016) no.~3, 031043},
\href{http://arxiv.org/abs/1606.01893}{{\tt arXiv:1606.01893 [hep-th]}}.

\bibitem{Seiberg:2016gmd}
N.~Seiberg, T.~Senthil, C.~Wang, and E.~Witten, ``{A Duality Web in 2+1
  Dimensions and Condensed Matter Physics},''
  \href{http://dx.doi.org/10.1016/j.aop.2016.08.007}{{\em Annals Phys.} {\bf
  374} (2016)  395--433},
\href{http://arxiv.org/abs/1606.01989}{{\tt arXiv:1606.01989 [hep-th]}}.

\bibitem{Dasgupta:1981zz}
C.~Dasgupta and B.~I. Halperin, ``{Phase Transition in a Lattice Model of
  Superconductivity},''
\href{http://dx.doi.org/10.1103/PhysRevLett.47.1556}{{\em Phys. Rev. Lett.}
  {\bf 47} (1981)  1556--1560}.

\bibitem{Peskin:1977kp}
M.~E. Peskin, ``{Mandelstam 't Hooft Duality in Abelian Lattice Models},''
\href{http://dx.doi.org/10.1016/0003-4916(78)90252-X}{{\em Annals Phys.} {\bf
  113} (1978)  122}.

\bibitem{Lee:1989fw}
D.~H. Lee and M.~P.~A. Fisher, ``{Anyon superconductivity and the fractional
  quantum Hall effect},''
\href{http://dx.doi.org/10.1103/PhysRevLett.63.903}{{\em Phys. Rev. Lett.} {\bf
  63} (1989)  903--906}.

\bibitem{Marino:1987tk}
E.~C. Marino, ``{Quantum Theory of Nonlocal Vortex Fields},''
\href{http://dx.doi.org/10.1103/PhysRevD.38.3194}{{\em Phys. Rev.} {\bf D38}
  (1988)  3194}.

\bibitem{Marino:1992uu}
E.~C. Marino, ``{Duality, quantum vortices and anyons in
  Maxwell-Chern-Simons-Higgs theories},''
  \href{http://dx.doi.org/10.1006/aphy.1993.1046}{{\em Annals Phys.} {\bf 224}
  (1993)  225--274},
\href{http://arxiv.org/abs/hep-th/9208062}{{\tt arXiv:hep-th/9208062
  [hep-th]}}.

\bibitem{Burgess:2000kj}
C.~P. Burgess and B.~P. Dolan, ``{Particle vortex duality and the modular
  group: Applications to the quantum Hall effect and other 2-D systems},''
  \href{http://dx.doi.org/10.1103/PhysRevB.63.155309}{{\em Phys. Rev.} {\bf
  B63} (2001)  155309},
\href{http://arxiv.org/abs/hep-th/0010246}{{\tt arXiv:hep-th/0010246
  [hep-th]}}.

\bibitem{Murugan:2014sfa}
J.~Murugan, H.~Nastase, N.~Rughoonauth, and J.~P. Shock, ``{Particle-vortex and
  Maxwell duality in the $AdS_4\times \mathbb{CP}^3$/ABJM correspondence},''
  \href{http://dx.doi.org/10.1007/JHEP10(2014)051}{{\em JHEP} {\bf 10} (2014)
  51},
\href{http://arxiv.org/abs/1404.5926}{{\tt arXiv:1404.5926 [hep-th]}}.

\bibitem{Ramos:2005yy}
R.~O. Ramos, J.~F. Medeiros~Neto, D.~G. Barci, and C.~A. Linhares, ``{Abelian
  Higgs model effective potential in the presence of vortices},''
  \href{http://dx.doi.org/10.1103/PhysRevD.72.103524}{{\em Phys. Rev.} {\bf
  D72} (2005)  103524},
\href{http://arxiv.org/abs/hep-th/0506052}{{\tt arXiv:hep-th/0506052
  [hep-th]}}.

\bibitem{Ramos:2007hk}
R.~O. Ramos and J.~F. Medeiros~Neto, ``{Transition Point for Vortex
  Condensation in the Chern-Simons Higgs Model},''
  \href{http://dx.doi.org/10.1016/j.physletb.2008.07.097}{{\em Phys. Lett.}
  {\bf B666} (2008)  496--501},
\href{http://arxiv.org/abs/0711.0798}{{\tt arXiv:0711.0798 [hep-th]}}.

\bibitem{Maldacena:1997re}
J.~M. Maldacena, ``{The Large N limit of superconformal field theories and
  supergravity},'' {\em Adv.Theor.Math.Phys.} {\bf 2} (1998)  231--252,
\href{http://arxiv.org/abs/hep-th/9711200}{{\tt arXiv:hep-th/9711200
  [hep-th]}}.

\bibitem{Nastase:2015wjb}
H.~Nastase, {\em {Introduction to the ADS/CFT Correspondence}}.
\newblock Cambridge University Press, Cambridge,
2015.
\newblock

\bibitem{Ammon:2015wua}
M.~Ammon and J.~Erdmenger, {\em {Gauge/gravity duality}}.
\newblock Cambridge University Press, Cambridge,
2015.
\newblock

\bibitem{Nastase:2018cfe}
H.~Nastase, ``{Towards deriving the AdS/CFT correspondence},''
\href{http://arxiv.org/abs/1812.10347}{{\tt arXiv:1812.10347 [hep-th]}}.

\bibitem{Witten:2003ya}
E.~Witten, ``{SL(2,Z) action on three-dimensional conformal field theories with
  Abelian symmetry},''
\href{http://arxiv.org/abs/hep-th/0307041}{{\tt arXiv:hep-th/0307041
  [hep-th]}}.

\bibitem{Herzog:2007ij}
C.~P. Herzog, P.~Kovtun, S.~Sachdev, and D.~T. Son, ``{Quantum critical
  transport, duality, and M-theory},''
  \href{http://dx.doi.org/10.1103/PhysRevD.75.085020}{{\em Phys. Rev.} {\bf
  D75} (2007)  085020},
\href{http://arxiv.org/abs/hep-th/0701036}{{\tt arXiv:hep-th/0701036
  [hep-th]}}.

\bibitem{Aharony:2008ug}
O.~Aharony, O.~Bergman, D.~L. Jafferis, and J.~Maldacena, ``{N=6 superconformal
  Chern-Simons-matter theories, M2-branes and their gravity duals},''
  \href{http://dx.doi.org/10.1088/1126-6708/2008/10/091}{{\em JHEP} {\bf 0810}
  (2008)  091},
\href{http://arxiv.org/abs/0806.1218}{{\tt arXiv:0806.1218 [hep-th]}}.

\bibitem{LeGuillou:1996dv}
J.~C. Le~Guillou, E.~Moreno, C.~Nunez, and F.~A. Schaposnik, ``{NonAbelian
  bosonization in two-dimensions and three-dimensions},''
  \href{http://dx.doi.org/10.1016/S0550-3213(96)00676-1}{{\em Nucl. Phys.} {\bf
  B484} (1997)  682--696},
\href{http://arxiv.org/abs/hep-th/9609202}{{\tt arXiv:hep-th/9609202
  [hep-th]}}.

\bibitem{LeGuillou:1997zx}
J.~C. Le~Guillou, E.~Moreno, C.~Nunez, and F.~A. Schaposnik, ``{On
  three-dimensional bosonization},''
  \href{http://dx.doi.org/10.1016/S0370-2693(97)00857-5}{{\em Phys. Lett.} {\bf
  B409} (1997)  257--264},
\href{http://arxiv.org/abs/hep-th/9703048}{{\tt arXiv:hep-th/9703048
  [hep-th]}}.

\bibitem{Nastase:2017cxp}
H.~Nastase, {\em {String Theory Methods for Condensed Matter Physics}}.
\newblock Cambridge University Press,
2017.
\newblock

\bibitem{Iqbal:2008}
N.~Iqbal and H.~Liu, ``{Universality of the hydrodynamic limit in AdS/CFT and
  the membrane paradigm},''
  \href{http://dx.doi.org/10.1103/PhysRevD.79.025023}{{\em Phys.Rev.D} {\bf 79}
  (2009)  025023},
\href{http://arxiv.org/abs/0809.3808}{{\tt arXiv:0809.3808 [hep-th]}}.

\bibitem{Lopez-Arcos:2013uga}
C.~Lopez-Arcos, H.~Nastase, F.~Rojas, and J.~Murugan, ``{Conductivity in the
  gravity dual to massive ABJM and the membrane paradigm},''
  \href{http://dx.doi.org/10.1007/JHEP01(2014)036}{{\em JHEP} {\bf 01} (2014)
  036},
\href{http://arxiv.org/abs/1306.1263}{{\tt arXiv:1306.1263 [hep-th]}}.

\bibitem{Kovtun:2003wp}
P.~Kovtun, D.~T. Son, and A.~Starinets, ``{Holography and hydrodynamics:
  diffusion on stretched horizons},''
  \href{http://dx.doi.org/10.1088/1126-6708/2003/10/064}{{\em JHEP} {\bf 0310}
  (2003)  064},
\href{http://arxiv.org/abs/hep-th/0309213}{{\tt arXiv:hep-th/0309213
  [hep-th]}}.

\bibitem{Brigante:2007nu}
M.~Brigante, H.~Liu, R.~C. Myers, S.~Shenker, and S.~Yaida, ``{Viscosity Bound
  Violation in Higher Derivative Gravity},''
  \href{http://dx.doi.org/10.1103/PhysRevD.77.126006}{{\em Phys. Rev.} {\bf
  D77} (2008)  126006},
\href{http://arxiv.org/abs/0712.0805}{{\tt arXiv:0712.0805 [hep-th]}}.

\bibitem{Myers:2010pk}
R.~C. Myers, S.~Sachdev, and A.~Singh, ``{Holographic Quantum Critical
  Transport without Self-Duality},''
  \href{http://dx.doi.org/10.1103/PhysRevD.83.066017}{{\em Phys. Rev.} {\bf
  D83} (2011)  066017},
\href{http://arxiv.org/abs/1010.0443}{{\tt arXiv:1010.0443 [hep-th]}}.

\bibitem{Banks:2015wha}
E.~Banks, A.~Donos, and J.~P. Gauntlett, ``{Thermoelectric DC conductivities
  and Stokes flows on black hole horizons},''
  \href{http://dx.doi.org/10.1007/JHEP10(2015)103}{{\em JHEP} {\bf 10} (2015)
  103},
\href{http://arxiv.org/abs/1507.00234}{{\tt arXiv:1507.00234 [hep-th]}}.

\bibitem{Donos:2017mhp}
A.~Donos, J.~P. Gauntlett, T.~Griffin, N.~Lohitsiri, and L.~Melgar,
  ``{Holographic DC conductivity and Onsager relations},''
  \href{http://dx.doi.org/10.1007/JHEP07(2017)006}{{\em JHEP} {\bf 07} (2017)
  006},
\href{http://arxiv.org/abs/1704.05141}{{\tt arXiv:1704.05141 [hep-th]}}.

\bibitem{Fischler:2015kro}
W.~Fischler and S.~Kundu, ``{Membrane paradigm, gravitational $\Theta$-term and
  gauge/gravity duality},''
  \href{http://dx.doi.org/10.1007/JHEP04(2016)112}{{\em JHEP} {\bf 04} (2016)
  112},
\href{http://arxiv.org/abs/1512.01238}{{\tt arXiv:1512.01238 [hep-th]}}.

\bibitem{Parikh:1997ma}
M.~Parikh and F.~Wilczek, ``{An Action for black hole membranes},''
  \href{http://dx.doi.org/10.1103/PhysRevD.58.064011}{{\em Phys. Rev.} {\bf
  D58} (1998)  064011},
\href{http://arxiv.org/abs/gr-qc/9712077}{{\tt arXiv:gr-qc/9712077 [gr-qc]}}.

\bibitem{Deser:1984kw}
S.~Deser and R.~Jackiw, ``{'Selfduality' of Topologically Massive Gauge
  Theories},''
\href{http://dx.doi.org/10.1016/0370-2693(84)91833-1}{{\em Phys. Lett.} {\bf
  B139} (1984)  371--373}.

\bibitem{Townsend:1983xs}
P.~K. Townsend, K.~Pilch, and P.~van Nieuwenhuizen, ``{Selfduality in Odd
  Dimensions},'' \href{http://dx.doi.org/10.1016/0370-2693(84)91753-2,
  10.1016/0370-2693(84)92051-3}{{\em Phys. Lett.} {\bf B136} (1984)  38}.
[Addendum: Phys. Lett.B137,443(1984)].

\bibitem{Qi:2008ew}
X.-L. Qi, T.~Hughes, and S.-C. Zhang, ``{Topological Field Theory of
  Time-Reversal Invariant Insulators},''
  \href{http://dx.doi.org/10.1103/PhysRevB.78.195424}{{\em Phys. Rev.} {\bf
  B78} (2008)  195424},
\href{http://arxiv.org/abs/0802.3537}{{\tt arXiv:0802.3537
  [cond-mat.mes-hall]}}.

\bibitem{Son:2015xqa}
D.~T. Son, ``{Is the Composite Fermion a Dirac Particle?},''
  \href{http://dx.doi.org/10.1103/PhysRevX.5.031027}{{\em Phys. Rev.} {\bf X5}
  (2015) no.~3, 031027},
\href{http://arxiv.org/abs/1502.03446}{{\tt arXiv:1502.03446
  [cond-mat.mes-hall]}}.

\bibitem{Lippert:2014jma}
M.~Lippert, R.~Meyer, and A.~Taliotis, ``{A holographic model for the
  fractional quantum Hall effect},''
  \href{http://dx.doi.org/10.1007/JHEP01(2015)023}{{\em JHEP} {\bf 01} (2015)
  023},
\href{http://arxiv.org/abs/1409.1369}{{\tt arXiv:1409.1369 [hep-th]}}.

\bibitem{Price:1986yy}
R.~H. Price and K.~S. Thorne, ``{Membrane Viewpoint on Black Holes: Properties
  and Evolution of the Stretched Horizon},''
\href{http://dx.doi.org/10.1103/PhysRevD.33.915}{{\em Phys. Rev.} {\bf D33}
  (1986)  915--941}.

\bibitem{Eling:2009sj}
C.~Eling and Y.~Oz, ``{Relativistic CFT Hydrodynamics from the Membrane
  Paradigm},'' \href{http://dx.doi.org/10.1007/JHEP02(2010)069}{{\em JHEP} {\bf
  02} (2010)  069},
\href{http://arxiv.org/abs/0906.4999}{{\tt arXiv:0906.4999 [hep-th]}}.

\bibitem{Nastase:2015ljb}
H.~Nastase, ``{DBI scalar field theory for QGP hydrodynamics},''
  \href{http://dx.doi.org/10.1103/PhysRevD.94.025014}{{\em Phys. Rev.} {\bf
  D94} (2016) no.~2, 025014},
\href{http://arxiv.org/abs/1512.05257}{{\tt arXiv:1512.05257 [hep-th]}}.

\bibitem{Endlich:2010hf}
S.~Endlich, A.~Nicolis, R.~Rattazzi, and J.~Wang, ``{The Quantum mechanics of
  perfect fluids},'' \href{http://dx.doi.org/10.1007/JHEP04(2011)102}{{\em
  JHEP} {\bf 04} (2011)  102},
\href{http://arxiv.org/abs/1011.6396}{{\tt arXiv:1011.6396 [hep-th]}}.

\bibitem{Dubovsky:2011sj}
S.~Dubovsky, L.~Hui, A.~Nicolis, and D.~T. Son, ``{Effective field theory for
  hydrodynamics: thermodynamics, and the derivative expansion},''
  \href{http://dx.doi.org/10.1103/PhysRevD.85.085029}{{\em Phys. Rev.} {\bf
  D85} (2012)  085029},
\href{http://arxiv.org/abs/1107.0731}{{\tt arXiv:1107.0731 [hep-th]}}.

\bibitem{Berezhiani:2016dne}
L.~Berezhiani, J.~Khoury, and J.~Wang, ``{Universe without dark energy: Cosmic
  acceleration from dark matter-baryon interactions},''
  \href{http://dx.doi.org/10.1103/PhysRevD.95.123530}{{\em Phys. Rev.} {\bf
  D95} (2017) no.~12, 123530},
\href{http://arxiv.org/abs/1612.00453}{{\tt arXiv:1612.00453 [hep-th]}}.

\bibitem{Mukhi:2008ux}
S.~Mukhi and C.~Papageorgakis, ``{M2 to D2},''
  \href{http://dx.doi.org/10.1088/1126-6708/2008/05/085}{{\em JHEP} {\bf 05}
  (2008)  085},
\href{http://arxiv.org/abs/0803.3218}{{\tt arXiv:0803.3218 [hep-th]}}.

\end{thebibliography}\endgroup
\bibliographystyle{utphys}

\end{document}